\documentclass[10pt,journal,compsoc]{IEEEtran}
\ifCLASSOPTIONcompsoc
  \usepackage[nocompress]{cite}
\else
  \usepackage{cite}
\fi
\ifCLASSINFOpdf
\else
\fi
\usepackage[utf8]{inputenc}
\usepackage{color}
\usepackage{amssymb}
\usepackage{pifont}
\usepackage{multirow}

\newcommand{\hash}{\textsc{Trust}}

\usepackage{url}
\usepackage{flushend}
\usepackage{algorithm}
\usepackage{algorithmicx}
\usepackage{algpseudocode}

\usepackage{graphicx}
\usepackage{subfig}
\usepackage{amsmath}
\usepackage{wrapfig}
\usepackage{enumitem}  
\usepackage{anyfontsize}

\newcommand{\update}[1]{{\textcolor{black}{#1}}}
\newcommand{\final}[1]{{\textcolor{black}{#1}}}

\begin{document}
\title{ {\hash}: Triangle Counting Reloaded on GPUs}

\author{Santosh~Pandey,~Zhibin~Wang,~Sheng~Zhong,~Chen~Tian,~Bolong~Zheng,~Xiaoye~Li,~Lingda~Li,~Adolfy~Hoisie,~Caiwen~Ding,~Dong~Li,~Hang~Liu
  
\IEEEcompsocitemizethanks{
\IEEEcompsocthanksitem S. Pandey and H. Liu are with the Electrical and Computer Engineering, Stevens Institute of Technology.\protect\\
E-mail: {spande1, hang.liu} @stevens.edu.
\IEEEcompsocthanksitem Z. Wang, S. Zhong and C. Tian are with the State Key Laboratory for Novel Software Technology, Nanjing University, Nanjing, Jiangsu 210008,China.\protect\\
E-mail: {wzbwangzhibin, sheng.zhong}@gmail.com, tianchen@nju.edu.cn.
\IEEEcompsocthanksitem B. Zheng is with the Huazhong University of Science and Technology.\protect\\
E-mail: bolongzheng@hust.edu.cn.
\IEEEcompsocthanksitem X. Li is with the Computational Research Division, Lawrence Berkeley National Laboratory.\protect\\
E-mail: xsli@lbl.gov.
\IEEEcompsocthanksitem L. Li and A. Hoisie are with the Brookhaven National Laboratory.\protect\\
E-mail: {lli, ahoisie} @bnl.gov.
\IEEEcompsocthanksitem C. Ding is with the Department of Computer Science \& Engineering, University of Connecticut.\protect\\
E-mail: caiwen.ding@uconn.edu.
\IEEEcompsocthanksitem D. Li is with the Department of Electrical Engineering and Computer Science, University of California, Merced.\protect\\
E-mail: dli35@ucmerced.edu.
}
\thanks{
Santosh Pandey and Zhibin Wang contributed to the work equally. \\
(Corresponding author: Hang Liu.)
}
}




\IEEEtitleabstractindextext{%
\begin{abstract}
    Triangle counting is a building block for a wide range of graph applications.
    Traditional wisdom suggests that i) ha{s}hing is not suitable for triangle counting, ii) edge-centric triangle counting beats ver{t}ex-centric design, and iii) communication-free and workload balanced graph partitioning is a grand challenge for triangle counting.
    On the contrary, we advocate that i) hashing can help the key operations for scalable triangle counting on Graphics Processing Units (GPUs), i.e., list intersection and graph partitioning, ii) vertex-centric design reduces both hash table construction cost and memory consumption, which is limited on GPUs. In addition, iii) we exploit graph and workload collaborative, and hashing-based 2D partitioning to scale vertex-centric triangle counting over 1,000 GPUs with sustained scalability. In this work, we present {\hash} which performs \underline{tr}iangle co\underline{u}nting with the ha\underline{s}h operation and ver\underline{t}ex-centric mechanism at the core.
    To the best of our knowledge, {\hash} is the first work that achieves over \textit{one trillion} Traversed Edges Per Second (TEPS) rate for triangle counting. 

\end{abstract}

}

\maketitle
\IEEEdisplaynontitleabstractindextext
\IEEEpeerreviewmaketitle

\IEEEraisesectionheading{\section{Introduction}\label{sec:introduction}}

\noindent
The number of triangles (i.e., three-vertex clique) is a key metric to extract insights for a wide range of graph applications, such as, anomaly detection~\cite{noble2003graph,akoglu2015graph},
community detection~\cite{prat2014high,prat2016put,zhang2016efficient,rezvani2018efficient},
and robustness analysis~\cite{dekker2004network}. For more thorough studies about the applications surrounding triangle counting, we refer the readers to recent surveys~\cite{schank2007algorithmic,chu2011triangle,al2018triangle}.
Further, triangle counting is also a basic primitive for an array of graph algorithms, e.g., clustering coefficient~\cite{watts1998collective}, k-truss~\cite{wang2012truss,date2017collaborative,smith2017truss},
and transitivity ratio calculation~\cite{wasserman1994social}. Ultimately, the significance of triangle counting is pronounced by the GraphChallenge competition~\cite{graphchallenge}, where 
participants
are ranked by how fast they perform triangle counting on a collection of graph datasets.



\update{Recent years have witnessed a surge of projects in triangle counting.}
Briefly, triangle counting efforts fall into three categories, that is, list intersection, matrix-multiplication, and subgraph matching. 
List intersection further encompasses two system implementation methods, i.e., edge- and vertex- centric options. In terms of how to intersect the lists, one can exploit merge-path, binary-search, and hashing-based algorithms. Note, the bitmap is an extreme case of hashing where the number of buckets equals to the number of vertices. All the details about these methods are thoroughly discussed in Section~\ref{sec:background}. 

The efforts of seeking suitable hardware platforms to accelerate triangle counting has also gain momentum. Popular attempts include multi-core CPUs~\cite{shun2015multicore,pearce2017triangle,elenberg2015beyond,arifuzzaman2015fast},
many-core GPUs~\cite{hu2018tricore,hu2017trix,hu2018high,hoang2019disttc,pandey2019h}, 
and external memory devices~\cite{hu2013massive,giechaskiel2015pdtl,cui2018efficient,hu2014efficient,hu2016efficient}.
Of all these platforms, GPUs are particularly tempting for the following reasons. 
First and foremost, GPUs come with unprecedented computing and data delivering capabilities. Using recent NVIDIA Tesla V100~\cite{nvidia2017v100} GPU as an example, it provides 80 streaming multiprocessor (SM) and 64 FP32 cores/SM, which can reach 15.7 TFLOPS peak performance. Along with High Bandwidth Memory (HBM2) on the device, this GPU can retain 900 GB/s memory bandwidth. 
The massive parallelism and fast memory support are well suited for triangle counting. 
Second, GPUs are equipped with configurable on-chip shared memory where users can store frequently accessed data structures. As we will discuss shortly, shared memory can significantly improve the efficiency of triangle counting.
Last but not least, GPUs feature a hierarchical thread organization, e.g., thread, warp, and Cooperative Thread Array (CTA), which fits graphs that come with inherent workload imbalance across various vertices. 


\subsection{Related Work and Challenges}
\label{subsec:pitfall}

Reviewing the recent literatures centering around triangle counting,
we arrive at the following challenges faced by vertex-centric hashing-based triangle counting, along with brief discussions about our resolutions.  



\textit{Challenge 1}. 
The hashing-based list intersection is not suitable for triangle counting resulting from the concern of collision. Particularly, hashing-based intersection can count triangles as hashing puts identical elements into the same bucket. However, with limited buckets, hashing also puts different elements into the same bucket, known as collisions. 
To lower the collision cost,~\cite{shun2015multicore} allocates a gigantic memory space that is $l\times$ the original graph size. Afterwards, each vertex $u$ would take $l \times d(u)$, i.e., degree of $u$, space from the gigantic memory space to build $u$'s hash table. Empirically, $l$ could be 2 - 4 if we want the cost of the collision to be low. Given this design needs a large memory space for hash table, ~\cite{shun2015multicore} observes high cache misses for hashing-based designs and thus claims merge-path based method is better for triangle counting on CPUs. 
Later,~\cite{che2019accelerating,bisson2017high,bisson2018update}
use bitmap to represent the hash bucket which still suffers from high cache pressure. 
We also notice that \update{Yacsar et al.~\cite{yacsar2018fast,yacsar2019linear,wolf2017fast,acer2019scalable}}
switch between dense and sparse representations of a hash table in their matrix-multiplication effort which is, however, complex in nature. 


In this paper, by reordering the graph and adjusting GPU hardware resources with respect to the vertex degree, we turn collisions into a tolerable issue. 
Further, we fully unleash the potential of hashing, that is, using hashing for not only intersection, but also rapidly distributing workload across many-threads in one GPU, as well as across GPUs. 



\textit{Challenge 2}. Vertex-centric triangle counting is worse than the edge-centric counterpart on GPUs due to more severe workload imbalance issues~\cite{hu2018tricore,hu2017trix,hu2018high,pandey2019h}. 
Particularly, vertex-centric design~\cite{shun2015multicore,yacsar2018fast,yacsar2019linear,wolf2017fast,che2019accelerating} 
iterates through each vertex, loads the 1- and 2- hop neighbors, and intersects them to arrive at the triangles. Edge-centric design~\cite{hu2018tricore,hu2017trix,hu2018high,pandey2019h}
does that for each edge thus only 1-hop neighbors are needed. As a result, workload imbalance would arise from both inter- and intra- vertex aspects in vertex-centric design while the edge-centric counterpart only experiences workload imbalance across edges. 
Mathematically,
the time complexity of vertex-centric design is $\mathcal{O}(d(u)+\sum_{v\in N(u)}d(v))$ for vertex $u$ while that of edge-centric design is merely $\mathcal{O}(d(u)+d(v))$ between $u$ and $v$, where $d(u), d(v)$ and $N(u)$ are the degrees of $u$ and $v$, and the neighbor list of $u$, respectively. Hence, the workload difference between vertices is often higher than that of edges.
In terms of intra-vertex imbalance, 
for each vertex $u$, we need to intersect $u$'s neighbor list with all its 2-hop neighbor lists, where the workloads of different 2-hop neighbor lists are also likely to be dissimilar. 
\update{Note, both vertex-centric and edge-centric designs perform accurate triangle counting and result in the same number of triangles}.

While the vertex-centric design comes with the concern of imbalance, it also exhibits unique advantages. 
First, the vertex-centric design avoids the need of the graph in edge list format, which saves $\frac{2}{3}$ of space and data movement traffic~\cite{hu2018tricore}. Second, for hashing-based intersection, the vertex-centric design largely reduces the cost of constructing the hash table compared to the edge-centric method~\cite{pandey2019h}. Furthermore, we find that the innate GPU thread and memory hierarchy is a great remedy for workload imbalance. 

{\textit{Challenge 3}. The vertex-centric design makes distributed triangle counting a grand challenge stemming from the hardship of achieving communication free and workload balanced graph partitioning.}
As graphs continue to grow, a single machine (or device) will eventually fail to accommodate a large graph in the memory. 
As a result, researchers rely upon either external memory options~\cite{kumar2016g,hu2013massive,chu2011triangle,pearce2013scaling}
or distributed settings~\cite{hu2017trix,hu2018tricore}
to resolve this problem. 
In order to achieve better performance, both designs need communication free and workload balanced graph partitioning.
However, even for edge-centric design which only requires 1-hop neighbor lists, achieving both goals is challenging, which is evident both theoretically~\cite{hu2017trix,arifuzzaman2013patric} 
and practically~\cite{giechaskiel2015pdtl,hu2017trix,hu2018high,elenberg2015beyond,zhang2019litete}. 
The vertex-centric design requires 2-hop neighbors, which further exacerbates the imbalance and communication problems. 

In this work, we separate the goal of achieving communication free and workload balance during graph partitioning. For the first goal, we propose a 2D graph partitioning algorithm, that partitions the 1-hop neighbors and uses the 1-hop neighbor partitions to build the 2-hop ones so that the vertex range partitions of 1-hop neighbors are the same as the 2-hop ones. The workload balancing goal is achieved by hashing-based partitioning over our reordered graphs.
 And we further partition the workloads in order to scale {\hash} up to 1,000 GPUs. 

\subsection{Contributions}
This paper designs and implements a vertex-centric hashing-based triangle counting system on GPUs that can achieve beyond the trillion TEPS performance on random, rMat, and 3Dgrid graph datasets.  
Particularly, this work not only reveals and leverages the unique advantages of hashing and vertex-centric designs for scalable triangle counting on GPUs but also carefully designs optimizations to overcome the key challenges faced by the vertex-centric hashing method. 
In summary, this work makes the following contributions.

First, vertex-centric hashing presents great potentials for GPU-based triangle counting. 
In spite of collision concern, hashing-based intersection exhibits advantageous features over both merge-path~\cite{green2014fast} and binary-search~\cite{hu2018tricore} based counterparts~\cite{DBLP:conf/alenex/2014,DBLP:conf/alenex/OrtmannB14}.
Particularly, merge-path suffers from workload partitioning hardship, while hashing does not. Binary-search experiences high time complexity at $\mathcal{O}(logN)$, and hashing lowers that cost to $\mathcal{O}(1)$. \update{Furthermore, binary-search requires random access to the binary tree, while our interleaved hash table layout and linear search enjoys coalesced memory access.} 
For vertex- vs. edge- centric design comparison, vertex-cenric design only needs the graph in adjacency list format while the edge-centric design requires both edge list and adjacency list formats of a graph. 
Putting hashing and vertex-centric designs together, {\hash} avoids repeated hash table construction in edge-centric design~\cite{pandey2019h}.
On average, the vertex-centric design reduces the hash table construction time by 92$\times$.
When deployed on GPUs, we interleave the entries from all buckets and exploit GPU shared memory to lower the hash table lookup cost. 



\begin{table*}[htbp]
	\renewcommand\arraystretch{1}
	\scalebox{0.85}{
	\begin{tabular}{|l|c|c|c|c|c|c|c|}
		\hline
\multirow{3}{*}{} & \multicolumn{5}{c|}{\textbf{Intersection}} & \multirow{3}{*}{\textbf{Matrix-multiplication}} & \multirow{3}{*}{\textbf{Subgraph matching}} \\ \cline{2-6}
 & \multirow{2}{*}{\textbf{Binary-search}} & \multirow{2}{*}{\textbf{Merge-path}} & \multirow{2}{*}{\textbf{Bitmap}} & \multicolumn{2}{c|}{\textbf{Hashing}} &  &  \\ \cline{5-6}
		&  &  &  &   \textbf{Vertex-centric} & \textbf{Edge-centric} & &\\ \hline
		\textbf{CPU}  & ~\cite{date2017collaborative} & ~\cite{shun2015multicore,giechaskiel2015pdtl,hu2016efficient} & \cite{hu2013massive,hu2014efficient} & ~\cite{shun2015multicore} & & ~\cite{ azad2015parallel, Low2017FirstLL, acer2019scalable,yacsar2018fast,yacsar2019linear,wolf2017fast} & \cite{bi2016efficient,lai2015scalable}\\ \hline
		\textbf{GPU}  & ~\cite{hu2018tricore,hu2017trix,hu2018high,hu2019triangle, hoang2019disttc,date2017collaborative} & ~\cite{green2014fast} & ~\cite{bisson2018update,bisson2017high,bisson2017static} & \textbf{\hash} & ~\cite{pandey2019h}&  ~\cite{yacsar2019linear, wang2016comparative} & \cite{wang2016comparative}\\ \hline
	\end{tabular}
	}
	\caption{\update{Closely related projects for {\hash}.}}
	\label{tab:related}
	\vspace{-0.1in}
\end{table*}

Second, admittedly, vertex-centric hashing also comes with drawbacks, i.e., collisions and workload imbalance, which require optimizations. 
Towards collision reduction, we propose a graph reordering technique that reorders the vertex IDs of a graph.
Since optimal reordering is NP-complete, we find two effective heuristics. The intuition behind these heuristics is that we should prioritize the high-degree vertices and their neighbors when lowering the collisions. This approach enhances the performance by {up to 75\%}. For intra-vertex workload imbalance, we introduce a virtual combination method to virtually combine the 2-hop neighbors in order to ultimately balance the intra-vertex workload. This yields, on average, {50\%} speedup across all graphs. Taken collision and inter-vertex workload imbalance together, we introduce degree-aware resources allocation mechanisms that give large degree vertices more hash buckets, shared memory, and threads. This design yields, on average, {7$\times$} speedup across all the graphs. 

Third, we introduce graph and workload collaborative, hashing-based 2D partitioning scheme to scale triangle counting beyond 1,000 GPUs. 
Particularly, we use hashing, instead of vertex range, to partition the graph into 2D (i.e., partition both source and destination vertices) so that each partition comes with similar amounts of workload, thanks to our graph reordering method. Subsequently, for each 1-hop neighbor partition used for hash table construction, we use the 1-hop neighbor partitions to build up the 2-hop neighbor partitions because our hashing-based 2D partition ensures the source and destination vertices are evenly partitioned. The partitioning approach is detailed in Figure~\ref{fig:Graph partition}(b). Since different 2-hop partitions can enumerate the triangles independently,
we further introduce workload partitioning, which distributes various 2-hop neighbor partitions across more GPUs. Taken together, our graph and workload collaborative partitioning can saturate 1,024 GPUs with merely 64 graph partitions.
This design is not only space and workload balanced but also communication free. Particularly, for extremely large graphs, we achieve 1.9$\times$ speedup from 512 to 1,024 GPUs and beyond 600$\times$ speedup for medium graphs from 1 to 1,024 GPUs. 

\vspace{-0.1in}
\subsection{Paper Organization}
The rest of this paper is organized as follows. Section~\ref{sec:background} presents the background. 
Section~\ref{sec:hash} describes the novel {\hash} designs. 
Section~\ref{sec:opt} presents the optimization techniques for hash collision and workload imbalance.
Section~\ref{sec:scale} presents our workload and graph collaborative partition methods.
Section~\ref{sec:eval} evaluates the performance of {\hash} and Section~\ref{sec:conclusion} concludes.

\section{Background}
\label{sec:background}

\subsection{Notation and Terminology}

Let $G (V, E)$ be an undirected and unweighted graph, $V$ and $E$ be the vertex and edge sets of $G$, respectively. 
Graphs are often stored in the array style data structures, among which edge list and Compressed Sparse Row (CSR) formats are the mainstream options.
Particularly, an edge list is a collection of all the edge tuples in $G$, where each tuple $(u, v)$ is an edge from $u$ to $v$ in $G$. CSR format uses two arrays, i.e., begin position and adjacency list. The adjacency list is a concatenation of the out neighbor lists of all vertices, and the begin position specifies the starting position of the neighbor list of each vertex.

\vspace{-0.1in}
\subsection{Triangle Counting {Algorithms}}

This section describes the mainstream triangle counting algorithms, i.e., intersection and other alternatives - matrix-multiplication and subgraph matching based methods. \update{Table~\ref{tab:related} categorizes these closely related projects.}



\textbf{Intersection based approach} encompasses three algorithm options, i.e., merge-path, binary-search, and hashing, which could be implemented in either vertex-centric or edge-centric fashion. 
\textit{Merge-path based intersection} uses two pointers to scan through two lists from beginning to end in order to find the intersection between them. During scanning, the pointer that points to a smaller value will be increased. A triangle is enumerated if both pointers increase (i.e., they point to the same vertex). \update{\cite{shun2015multicore,giechaskiel2015pdtl,hu2016efficient} observe that merge-path suits CPU based triangle counting due to lower time complexity compared with binary-search and higher cache hit rate compared with hashing.}
\textit{Binary-search based intersection} organizes the longer list as a binary tree, and uses the shorter list as search keys. For each search key, it descends through the binary-search tree in order to find the equal entry, which is a triangle. \update{Hu et al.~\cite{hu2018tricore,hu2017trix,hu2018high} indicate that edge-centric binary-search fits GPU based triangle counting because of higher parallelism and more balanced workloads.}
\textit{Hashing-based intersection} constructs a hash table for one list, then uses the other list as search keys to find the common elements in the hash table. Particularly,~\cite{shun2015multicore} only allows one element in each hash bucket of the hash table, which is also referred to as open addressing. When collision surfaces, this method uses linear probing mechanism. To avoid the high cost of linear probing, this method creates many hash buckets in the hash table, leading to overwhelming space consumption. 
\textit{Bitmap}
can be thought of as a hash table with $|V|$ {buckets}, which eliminates collision but consumes significantly more memory. Bisson et al.~\cite{bisson2018update,bisson2017static,bisson2017high} also perform vertex-centric GPU-based triangle counting. However, these projects use bitmaps to implement hash tables, which suffer from high memory consumption and are hence only suitable for small graphs. Several triangle counting projects~\cite{hu2013massive,hu2014efficient} also explore the bitmap option since they rely upon large external memory storage for triangle counting.

\begin{figure*}[t]
	\centering
	\includegraphics[width=.95\textwidth]{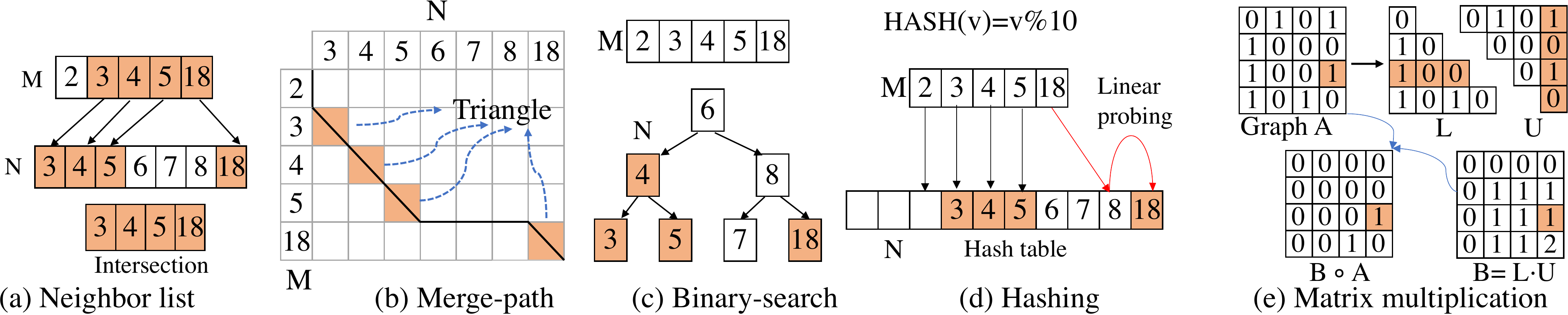}
	\caption{Four triangle counting methods. (a) M and N are the neighbor lists used by (b) - (d) which are merge-path, binary-search, and hashing-based intersection. And (e) uses matrix-multiplication to perform triangle counting for graph A.
	}
	\label{intersection}
	\vspace{-.15in}
\end{figure*}

Figure~\ref{intersection} explains how the aforementioned three intersection algorithms work on two lists M and N. As shown in Figure~\ref{intersection}(b), merge-path uses the vertical and horizontal pointers to scan through these two lists. Since the first element in M, i.e., 2, is smaller than that of N, the vertical pointer is increased. Further, because both M and N have 3 as their elements, both pointers are increased, and one triangle is enumerated. Similarly, we can enumerate all the triangles. In binary-search based method, as shown in Figure~\ref{intersection}(c), we use each element of M as the search key to search against the binary tree of N. For element 3 of M, the search keeps descend on the left side of N in order to find all the triangles. Figure~\ref{intersection}(d) depicts the hashing-based solution in~\cite{shun2015multicore}. This method first constructs a hash table for longer list N. Since $\textsc{hash}(18)=8$, element 18 first searches the index 8 in hash table, then linear probes to the next element which is 18, where a triangle is identified.

Existing intersection based approaches often exploit graph orientation to reduce the number of edges in the graph by half in order to reduce redundant work~\cite{shun2015multicore,pearce2017triangle}. For a pair of undirected edges, \textit{rank-by-degree} method, a representative graph orientation approach, removes the edge whose source degree is larger than the destination degree and preserves the remaining edge.

\textbf{Matrix-multiplication based approach}
decomposes the adjacency matrix (i.e., A) of the graph into lower and upper triangular matrices $\mathrm{L}$ and $\mathrm{U}$, respectively, as shown in Figure~\ref{intersection}(e).
Then it performs $\mathrm{B} = \mathrm{L}\cdot \mathrm{U}$, which counts the number of wedges. 
Further, the element-wise multiplication (i.e., Hadamard Product) of $\mathrm{A}$ and $\mathrm{B}$ determines whether the wedge is closed. Finally, we summarize the number of non-zero elements in the resultant matrix. 
Since each edge is counted by both vertices, the final sum is divided by 2 to get the exact count of triangles. Using Figure~\ref{intersection}(e) as an example, $\mathrm{L}[2][] = [1, 0, 0, 0]$ multiplying $\mathrm{U}[][3] = [1, 0, 1, 0]$ arrives at 1, i.e., $\mathrm{B}[2][3] = 1$, it means there is a wedge between $(2, 0)$ and $(0, 3)$. Afterwards, element-wise product between $\mathrm{A}[2][3]$ and $\mathrm{B}[2][3]$ can confirm whether there is an edge which closes the wedge. Yacsar et al.~\cite{yacsar2018fast,yacsar2019linear,wolf2017fast,acer2019scalable}, which leverage KokkosKernels linear algebra library~\cite{KokkosKernels} to count triangles, belong to this genre.

\textbf{Subgraph matching based approach}
searches for all occurrences of a query graph in a data graph. Triangle counting regards the triangle as that query graph.~\cite{wang2016comparative} implements a two-step subgraph matching approach for counting the number of triangles on undirected labeled graphs. First, the query graph - triangle in this case - is factored into a tree and the non-tree edges. 
Afterward, one finds all the vertices from the data graph that matches the root of the query tree using degree-based filtering. Subsequently, one traverses the query tree as well as the data graph from the candidates of the root with the matching rule. Finally, one joins the tree candidates and non-tree edge candidates to arrive at all the triangles in the data graph. 

\vspace{-0.1in}
\subsection{Approximate Triangle Counting}
{Since triangle counting in extremely large graphs is computationally expensive, some researchers also explore approximate triangle counting algorithms to reduce the runtime~\cite{tsourakakis2009doulion,pagh2012colorful,shun2015multicore,arifuzzaman2013patric,wang2019rept,wang2017approximately,rahman2013approximate,mawhirter2018approxg,wu2016counting,avron2010counting,tsourakakis2011spectral,tsourakakis2011counting}.
Among them, \cite{tsourakakis2009doulion,rahman2013approximate} {estimate the number of triangles by sampling the edges, and only counting triangles for the sampled edges.}
\cite{pagh2012colorful,shun2015multicore} first color the vertices, then keep the edges that connect two same-colored vertices. Further, they count the triangles in the sampled subgraphs and estimate the total triangles in the graph. \cite{seshadhri2013triadic} approximates the triangle count by wedge sampling. \cite{wu2016counting} performs a detailed experiment to compare these different sampling approaches. In addition to graph sampling-based method, 
\cite{avron2010counting,tsourakakis2011spectral,tsourakakis2011counting} approximate the count of triangles based on spectral decomposition of the graph.} 

\subsection{Hardware Platforms for Counting Triangles}

In addition to only using either CPU or GPU to count triangles,~\cite{liu2018griffin,yacsar2019linear} use CPU and GPU together to accelerate the intersection computation. We also find out that~\cite{huang2018triangle} deploys triangle counting on FPGAs, which presents better energy efficiency. 

\subsection{Graph Partitioning Methods}
\label{sec: related of partition}

Value-range based partitioning, such as 1D~\cite{liu2015enterprise,pandey2020c} and 2D partitioning~\cite{bulucc2011parallel}, is one of the most popular approach for triangle counting. In this direction, 
~\cite{suri2011counting,park2013efficient} 
design a 2D graph partitioning based on MapReduce, but suffer from workload imbalance. 
\cite{hu2017trix,hu2018tricore}
balance the workload of 2D partitioning by a runtime workload stealing scheme. However, this introduces nontrivial overheads. 
\cite{yacsar2019linear}
deploys 2D partitioning on matrix-multiplication based triangle counting. Whereas, the workload imbalance problem still exists.
\cite{elenberg2015beyond,hoang2019disttc} 
distribute edges of graph among different machines and cache the vertices requiring communication among the machines during triangle counting. 
\update{METIS~\cite{karypis1998fast} is a well-known topology-aware graph partitioning approach that aims to make balanced vertex/edge yet with lower edge cuts. This method, however, would require inter-worker communication when counting triangles.
Recently, LiteTe~\cite{zhang2019litete} also attempts to use value-range based 2D partitioning for triangle counting, but, again, experiences workload imbalance.
}

\subsection{Graph Dataset}

\begin{table}[htbp]
    \renewcommand\arraystretch{1.2}
    \centering
    \scalebox{0.89}{
    \begin{tabular}{|c|c|r|r|r|}
        \hline
        \textbf{Dataset} & \textbf{Abbr.} & \multicolumn{1}{|c|}{\textbf{$|V|$}} & \multicolumn{1}{|c|}{\textbf{$|E|$}}  & \multicolumn{1}{|c|}{\textbf{\# Triangles}} \\ \hline
        3Dgrid & 3D    & 99,897,344  & 299,692,032  & 0  \\
        random & RA    & 100,000,000  & 999,999,892  & 1,221  \\ 
        rMat & RM    & 129,594,758  & 996,771,953  & 4,114,616  \\\hline
        Cit-Patents & CP    & 3,774,768  & 16,518,947  & 7,515,023  \\
        Friendster & FS    & 65,608,366  & 1,806,067,135  & 4,173,724,142  \\  
        gsh-2015-host & GH    & 68,660,142  & 1,502,666,069  & 520,901,310,734  \\ 
        it-2004 & IT    & 41,290,682  & 1,027,474,947  & 48,374,551,054  \\
        MAWI & MA   & 128,568,730  & 135,117,420  & 10  \\ 
        Orkut & OR    & 3,072,441  & 117,185,083  & 627,584,181  \\
        Twitter & TW    & 41,652,230  & 1,202,513,046  & 34,824,916,864  \\
        Wikipedia & WK    & 12,150,976  & 288,257,813  & 11,686,212,734  \\ \hline

        clueweb12 & CW &955,207,488 &37,372,179,311&1,995,295,290,765 \\
        uk-2014 & UK & 787,801,471 &	42,464,215,550 &	7,872,561,225,874 \\ \hline
    \end{tabular}%
    }
    \vspace{.05in}
    \caption{Graph datasets.}
    \label{tab:graph dataset}%
    \vspace{-0.1in}
\end{table}%

Table~\ref{tab:graph dataset} presents all the graphs that are used to evaluate {\hash}. Broadly, these datasets fall into three types, that is, synthetic graphs, regular real-world graphs and extremely large real-world graphs. Particularly, RA, RM, and 3D are generated by the Problem Based Benchmark Suite (PBBS)~\cite{shun2012brief}.
In the regular real-world graph categories, MA is the Internet traffic archive~\cite{mawi2012mawi}. CP, OR, and FS are from Stanford Network Analysis Project (SNAP) datasets~\cite{leskovec2016snap}. 
TW~\cite{kwak2010twitter} is the Twitter graph, and WK~\cite{konect:2017:wikipedia_link_en} is the English Wikipedia link graph.
The remaining are web graphs, i.e., IT and GH, as well as the extremely large real-world graphs, i.e., CW and UK from WebGraph~\cite{BoVWFI,BRSLLP,BMSB}. Our evaluation transforms the graph by following steps: i) removing the duplicate edges and self-loops; ii) transforming directed graphs to undirected graphs; and iii) removing orphan vertices. The size of the graph and number of triangles are also included in Table~\ref{tab:graph dataset}.

\section{{\hash}: Vertex-Centric hashing-based Triangle Counting}
\label{sec:hash}

The consensus from recent literatures~\cite{shun2015multicore,wang2016comparative,hu2018tricore} implies that merge-path is the ideal option for multi-core CPU while binary-search excels on many-core GPUs. Hashing is a poor option stemming from the fact that existing attempts often use large memory space to combat collisions, which ends up with overwhelming memory consumption and poor cache reuse.
Further, due to the concern of workload imbalance with vertex-centric design, the edge-centric design appears as the mainstream option for triangle counting~\cite{hu2018tricore}. 

This work advocates vertex-centric based hashing for triangle counting on GPUs because hashing can rapidly distribute workload across threads and GPUs, and vertex-centric approach reduces both the time for hash table construction and the memory space for graph datasets.


\begin{algorithm}[t]
	\caption{Vertex-centric hashing-based triangle counting.}
	\label{alg:Vertex-Centric Hashing for Triangle Counting}
	{\footnotesize
	\begin{algorithmic}[1]
		\ForAll{ $u\in V$ \textbf{in parallel}}\textcolor{blue}{\it \ \ //Main entry}
			\State $hashTable=\textsc{hashTableConstruction}(u.neighborList)$;
			\ForAll { $v\in u.neighborList$ \textbf{in parallel}}
				\State $count+=\textsc{intersection}({hashTable},v.neighborList)$;
			\EndFor
		\EndFor
\Statex
		
		\Function {hashTableConstruction}{$neighborList$}
			\For{ $i=0 \text{ to } bucketNumber-1$ \textbf{in parallel}}
				\State $hashTable(i).len=0 $
			\EndFor          
			\ForAll { $v\in neighborList$ \textbf{in parallel}}
				\State $i=\textsc{hash}(v)$;
				\State $len=\textbf{atomicAdd}(hashTable(i).len,1)$;
				\State $hashTable(i).element(len)=v$;
			\EndFor
			\State \Return $hashTable$;
		\EndFunction
\Statex
		\Function {intersection}{${hashTable},neighborList$}
			\ForAll { $w\in neighborList$ \textbf{in parallel}}
				\State $i=\textsc{hash}(w)$;
				\State $count+=\textsc{linearSearch}(hashTable(i),w)$;
			\EndFor
			\State \Return $count$;
		\EndFunction
\Statex
		\Function {linearSearch}{$bucket,w$}
			\For {$j$=0 to $bucket.len-1$}
				\If {$bucket.element(j)=w$}
					\State \Return 1;
				\EndIf
			\EndFor
		\State \Return 0;
		\EndFunction
\Statex
{
		\Function {hash}{$x$}
			\State \Return $x \% bucketNumber$;
		\EndFunction}
	\end{algorithmic}
	}
\end{algorithm}

\vspace{-0.1in}
\subsection{{\hash} Algorithm}
\label{subsec: trust algorith}

Algorithm~\ref{alg:Vertex-Centric Hashing for Triangle Counting} shows our vertex-centric hashing-based triangle counting algorithm, which mainly contains two steps: i) constructing hash table ($hashTable$) for the neighbor list ($neighborList$) of current vertex $u$, i.e., $u.neighborList$, ii) for each neighbor $v$ of $u$, searching whether $v$'s neighbors appear in the $hashTable$. 
While the majority of the variables in Algorithm~\ref{alg:Vertex-Centric Hashing for Triangle Counting} have self-explanatory names, we briefly describe how {\hash} handles collisions as follows.

Different from prior arts~\cite{shun2015multicore,bisson2017high,bisson2017static,bisson2018update},
{\hash} exploits a more efficient approach to handle collisions, that is, we allow a bucket to contain more than one element. Here, all the buckets are of the same size and allocated in a continuous memory region, which is slightly different from the classical dynamic chaining strategy.
In light of this design, each bucket $hashTable(i)$ has two fields, i.e., $hashTable(i).len$ and $hashTable(i).element$. The former field is the number of elements in bucket $i$. Here, $hashTable(i).len-1$ is also the number of collisions in this bucket. The latter field is an array that contains all the elements in bucket $i$, e.g., $hashTable(i).element(j)$ is the $j+1$-th element in this bucket. 
During $hashTable$ construction, we use atomic operation to allow concurrent write to $hashTable$, where \textbf{atomicAdd}($hashTable(i).len$, 1) returns the location for the new element.
During intersection, to determine whether $w$ is in a $hashTable$, we calculate $\textsc{hash}(w)$ which returns the bucket to search against.

{\hash} relies upon linear-search (line 25 of Algorithm~\ref{alg:Vertex-Centric Hashing for Triangle Counting}) to search within the bucket of interest, which counters the traditional wisdom that often prefers binary-search stemming from two reasons.
First, binary-search needs to sort all the elements in each hash bucket while linear-search does not. Second, since our $hashTable$ stores the elements of the same index across all buckets together (detailed in Section~\ref{subsec: hash layout}), linear-search enjoy \textit{coalesced global memory access} while binary-search does not.

Considering the memory cost of $hashTable$, we assign a fixed size of GPU global memory for each warp, subsequently reuse this space for each processing vertex. In implementation, we use 1,024 CTAs, each of which has 32 warps. Each $hashTable$ in a warp contains 32 buckets with the maximum collision number as 128. In this case, the total memory consumption for $hashTable$ is 512 MB.

\vspace{-0.05in}
\subsection{GPU-Friendly \textit{hashTable} Layout}
\label{subsec: hash layout}

\begin{figure}[t]
    \centering
    \includegraphics[width=.9\linewidth]{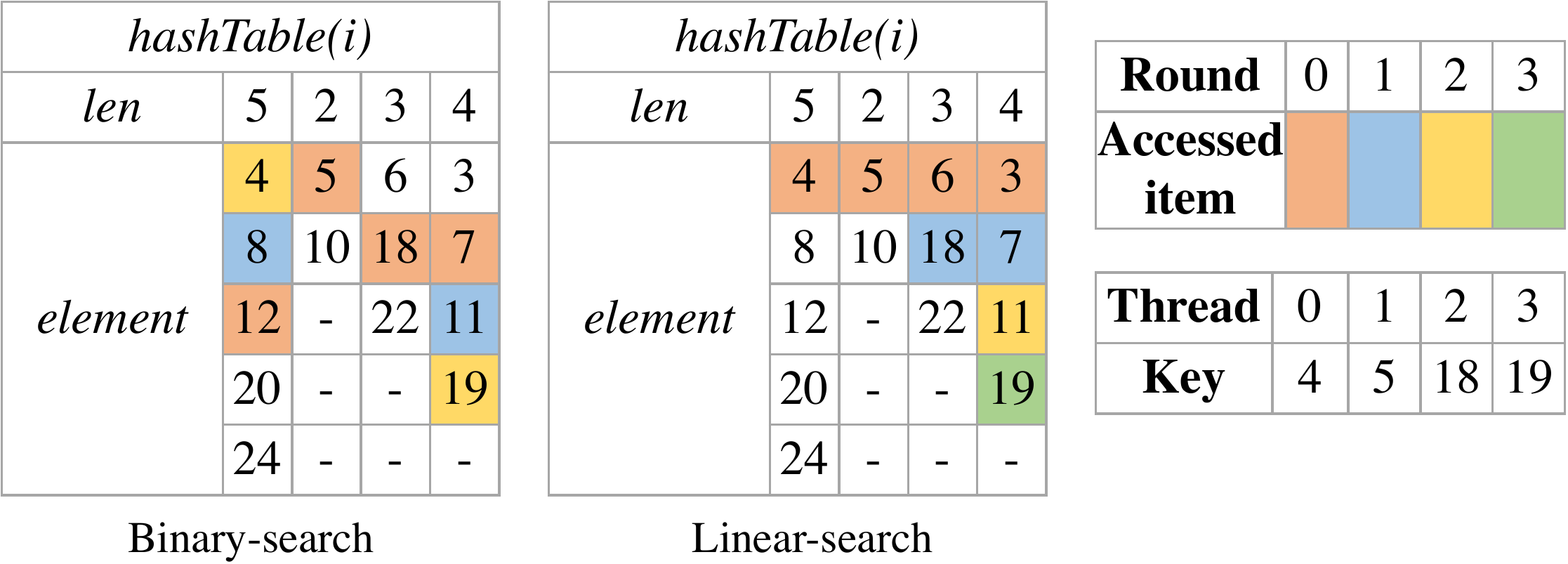}
	\vspace{-.05in}
    \caption{Linear- vs. binary- search for hash bucket search.
    \vspace{-.2in}
    }
	\label{fig:linearVsbinary}
\end{figure}

As shown in Figure~\ref{fig:linearVsbinary}, {\hash} further optimizes $hashTable$ layout including $hashTable(i).len$ and $hashTable(i).element$. First, we cache $hashTable(i).len$ in the shared memory. Second, we interleave the hash buckets of each $hashTable$ and cache the first few items of each bucket in shared memory.

We store $hashTable(i).len$ in shared memory because a significant number of buckets are empty, and storing $hashTable(i).len$ in shared memory avoids expensive global memory access. Further, $hashTable(i).len$ is frequently accessed during both construction and linear-search. During construction, $atomicAdd()$ in shared memory is much faster than in the global memory.

For $hashTable(i).element$, we optimize it in two ways. First, we store each level of a bucket consecutively instead of storing all the elements of a bucket consecutively so that consecutive threads access consecutive addresses. This leads to coalesced global memory access in linear-search. Using Figure~\ref{fig:linearVsbinary} as an example, the four hash buckets are \{4, 8, 12, 20, 24\}, \{5, 10\}, \{6, 18, 22\}, and \{3, 7, 11, 19\}. We store them as \{4, 5, 6, 3, 8, 10, 18, 7, 12, -, 22, 11, 20, -, -, 19, 24, -, -, -\} in memory. In this example, one GPU global memory access transaction can load four adjacent elements, which is one row in this particular case. During binary-search, the four threads accesses 5, 7, 12, and 18 in the first round, which leads to three global memory transactions. In contrast, linear-search accesses 4, 5, 6, and 3 in the first round, which is merely one global memory access transaction. Overall, in this example, linear-search performs four global memory access transactions while binary-search needs seven. 
Second, we store the first several elements of each bucket in the shared memory. Note, it is not always better to cache more elements in shared memory due to the occupancy concern~\cite{harris2007optimizing}. Further, recent GPU architectures, such as, V100, adopts a unified shared memory/L1 cache~\cite{nvidia2017v100}. Using more shared memory reduces the L1 cache size thus hurts the overall performance.

\begin{figure}[h]
    \centering
	\includegraphics[width=\linewidth]{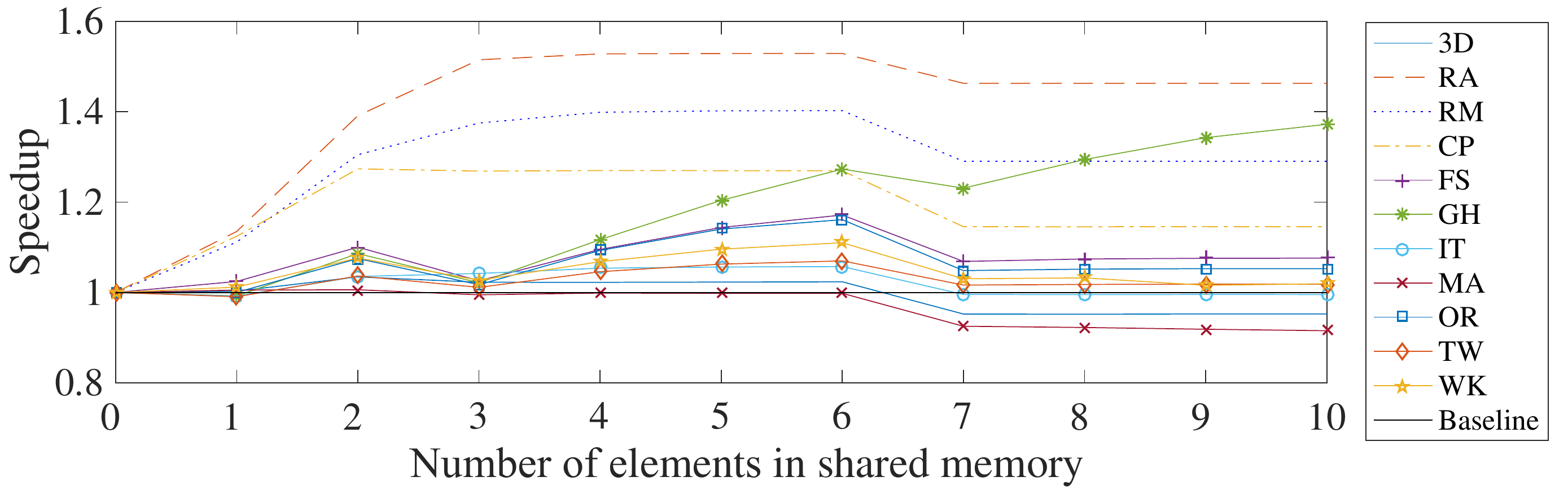}
    \caption{Hash memory optimizations. 
	\vspace{-0.1in}
	}
	\label{fig:hash mem opt}
\end{figure}

Figure~\ref{fig:hash mem opt} shows the performance of {\hash} with respect to the number of elements cached in each bucket. 
We observe that the performance climbs with the increase of cached elements for {RA, RM, CP, and GH} while the remaining graphs retain similar or worse performance. In this work, {\hash} caches 6 elements in shared memory for each bucket.

In an unlikely case, the $hashTable$ may reach the max collision threshold. In that case, linear probing is used to determine the next bucket for storing the neighbor. Consequently, during triangle counting, when a bucket is full, we need to perform linear-search in more than one bucket. Since linear probing is expensive, our optimizations (Section~\ref{sec:opt}) and partitioning schemes (Section~\ref{sec:scale}) are designed to avoid this. In our tests, the max collision across all graphs is often no more than 16 while our bucket size threshold is 128.

\vspace{-0.1in}
\subsection{Vertex-Centric Hashing}
\vspace{-0.05in}
\label{subsec:Vertex-centric hashing}

\begin{algorithm}[h]
	\caption{Edge-centric hashing-base triangle counting}
	\label{alg:edge-centric triange counting}
	{\footnotesize
	\begin{algorithmic}[1]
		\For {\textbf{all} $(u,v)\in E$ \textbf{in parallel}}
			\State $hashTable =\textsc{hashTableConstruction}(u.neighborList)$;
			\State $count+=\textsc{intersection}(hashTable,v.neighborList)$;
		\EndFor
	\end{algorithmic}
	}
\end{algorithm}

We observe hashing-based intersection favors the vertex-centric design despite that traditional efforts prefer the edge-centric design. The reason lies in that \textit{we need to construct $hashTable$ before intersection, and $hashTable$ construction time is also included in the total execution time}~{\cite{pandey2019h}}. Note, if that time is excluded, the comparison between {\hash} and other related works would be unfair. 
We further find that even if we were permitted to construct the $hashTable$ before counting triangles, $hashTable$ often consume significantly more memory than the $neighborList$ format, which is not suitable for GPUs that install limited memory space.

\begin{figure}[t]
	\centering
	\subfloat[Edge-centric design]{
		\includegraphics[width=0.49\linewidth]{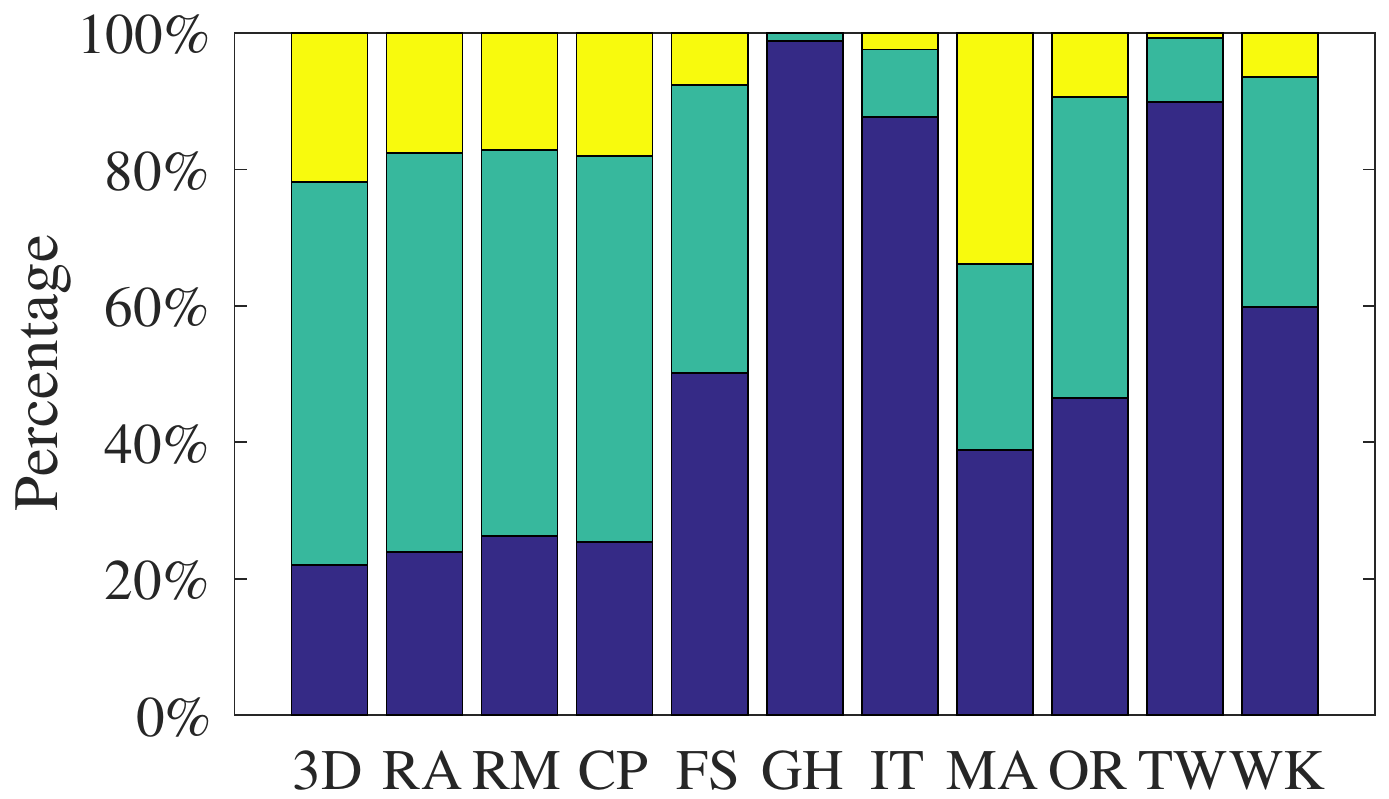}
	}
	\subfloat[Vertex-centric design]{
		\hspace{-.1in}\includegraphics[width=0.49\linewidth]{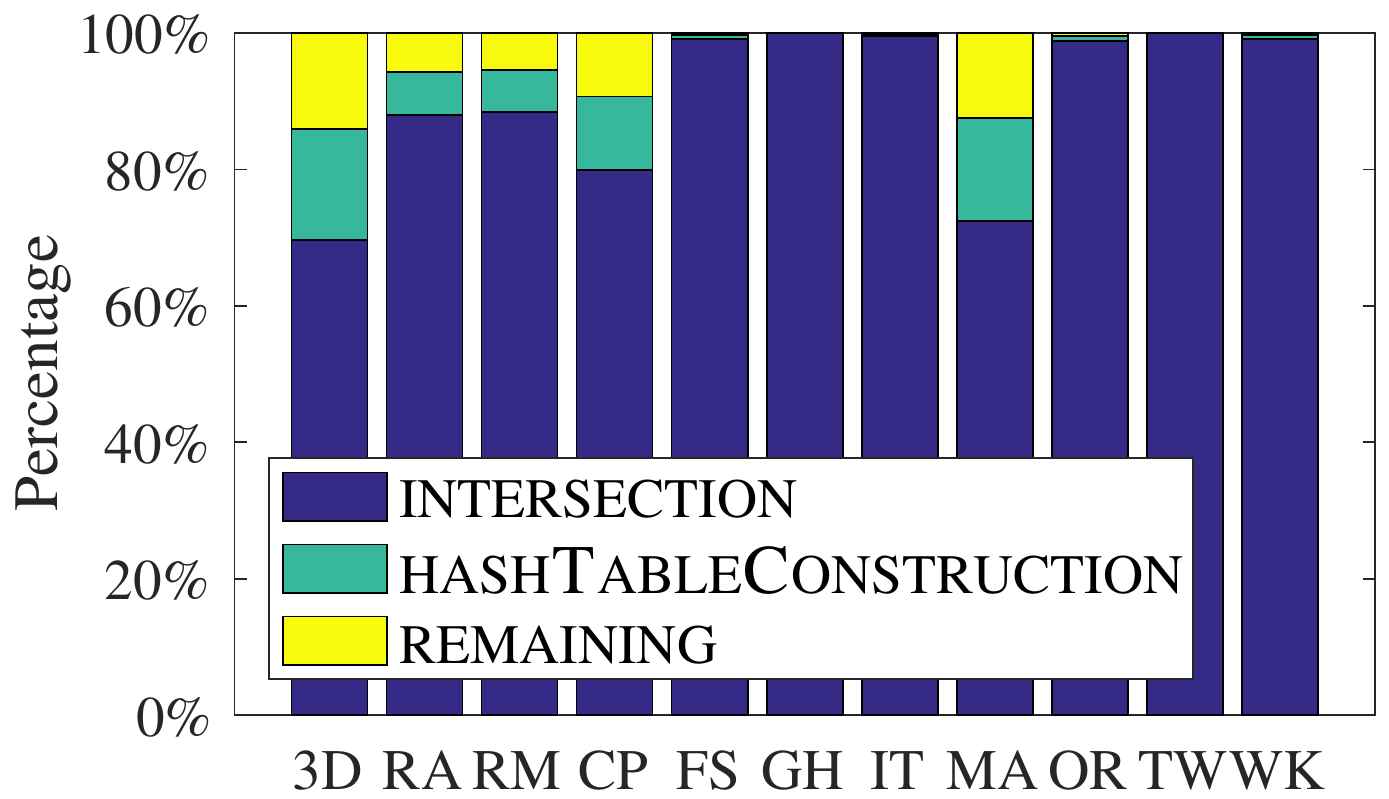}
	}
	
	\caption{Time consumption percentage of $hashTable$ construction, intersection, and the remaining for (a) edge-centric and (b) vertex-centric designs, respectively.
	\vspace{-.1in}
    }
	\label{fig:time cost between hash and intersection}
\end{figure}

Vertex-centric design consumes significantly shorter time than edge-centric design on $hashTable$ construction since vertex-centric option only constructs $hashTable$ once for each vertex, while the edge-centric counterpart needs to do that repeatedly. As shown in Algorithm~\ref{alg:edge-centric triange counting}, for each destination vertex $v$ of $u$, we need to construct the $hashTable$ for $u$, which is time consuming. As shown in Figure~\ref{fig:time cost between hash and intersection}, the time consumption ratio of $hashTable$ construction is 1\% - 57\% in edge-centric option~\cite{pandey2019h}. In contrast, our vertex-centric design reduces the $hashTable$ construction time ratio to 0.007\% - 16\%. When it comes to absolute time consumption, $hashTable$ construction time of vertex-centric design is reduced by 12.9$\times$ (RM) to 199.6$\times$ (GH), on average, 92$\times$ when compared to that of edge-centric design.

Vertex-centric hashing reduces the memory consumption for graph data.
Particularly, vertex-centric hashing does not require the edge list format of the graph which is needed by edge-centric counterpart. Note, edge list consumes about 2$\times$ memory compared with CSR format. Alternatively, TriCore~\cite{hu2018tricore} proposes to stream the edge list from CPU to GPU memory in order to reduce the memory consumption for the edge list. However, this design significantly affects the triangle counting performance as pointed by the recent study~\cite{hu2018high}. 


\vspace{-0.1in}
\section{Collision Reduction and Workload Balancing Optimizations}
\label{sec:opt}

Once the $hashTable$ construction time is significantly reduced by Section~\ref{sec:hash}, intersection becomes the bottleneck as shown in Figure~\ref{fig:time cost between hash and intersection}(b). This section optimizes the intersection through collision reduction and workload balancing. 

\vspace{-0.05in}
\subsection{Graph Reordering for Collision Reduction}\label{subsec:reorder}

According to Algorithm~\ref{alg:Vertex-Centric Hashing for Triangle Counting}, the cost of intersection can be formulated as Equation~(\ref{eq:cost}), assuming each 2-hop neighbor $w$ of $u$  needs to search through the entire bucket $hashTable_u(\textsc{hash}(w))$: \vspace{-0.1in}

\begin{align}\label{eq:cost}
	\vspace{-0.1in}
	\setlength{\abovedisplayskip}{0pt}
	\setlength{\belowdisplayskip}{0pt}
	 \sum_{u\in V}\sum_{v\in N(u)}\sum_{w\in N(v)}hashTable_u(\textsc{hash}(w)).len, 
	\vspace{0.1cm}
\end{align}
where $hashTable_u$, $N(u)$ and $N(v)$ represent the $hashTable$ for $u$, $u.neighborList$ and $v.neighborList$, respectively. 

Putting the analysis of Equation~(\ref{eq:cost}) in the GPU context, where a warp of threads work on 32 2-hop neighbors in \textit{Single Instruction Multiple Thread} fashion, the cost of {linear-search} is approximately decided by the max collision of all the buckets in a $hashTable$. Consequently, we arrive at the following estimation: 


\begin{align}\label{eq:cost_est}
	\vspace{-0.1in}
	\setlength{\abovedisplayskip}{0pt}
	\setlength{\belowdisplayskip}{0pt}
	 \phi=\sum_{u\in V}\underbrace{(\sum_{v\in N(u)}{degree(v)})}_\text{Collective\ degree of $u$}\cdot \underbrace{\text{max}(hashTable_u.len)}_\text{Max collision of $u$}. 
	\vspace{0.1cm}
\end{align}

Simply put, for each vertex $u$, the cost is proportional to the collective degrees of all neighbors of $u$, i.e., $\sum_{v\in N(u)}{degree(v)}$, as well as maximum collision of this $hashTable$ of $u$. 
Optimizing the order of the entire graph to arrive at the minimal cost for {Equation~(\ref{eq:cost_est})} is {NP-complete}, according to similar efforts for locality improvement~\cite{han2018speeding}.
Given the complex nature of this problem, we explore the following two heuristics to reduce the maximum hash collision guided by Equation~(\ref{eq:cost_est}). Particularly, since reordering does not affect the collective degree of $u$ in Equation~(\ref{eq:cost_est}), our reordering can only change the maximum collision. 
Note, these two techniques are separate and can not be used together.

\begin{itemize}
    \vspace{0.1cm}
    \item \textbf{Reordering by indegree} is guided by the fact that a vertex with higher indegree is more likely to appear in the $neighborList$ of other vertices. Consequently, this indegree method proposes to assign continuous IDs to vertices based upon their indegrees. In this way, large indegree vertices will have different hash values because their IDs are continuous. During $hashTable$ construction, these vertices are more likely to appear in the same $neighborList$ and less likely to be hashed into the same bucket, leading to a lower chance of maximum collision. For reordering, the vertices need to be sorted by their indegree. So, the time complexity is $\mathcal{O}(|V|log|V|)$.

    \vspace{0.15cm}
    \item \textbf{Reordering the neighbors of the largest collective outdegree first} is guided by the collective degree of $u$ in Equation~(\ref{eq:cost_est}). 
    Particularly, we observe that if we choose to minimize the maximum collision of the vertices with the largest collective degree, the cost $\phi$ will reduce: 
    i) This collective method sorts the vertices based upon their collective degrees. ii) For each $v\in u.neighborList$, if it does not have an assigned ID, we assign a new ID to it, where the new ID grows continuously. 
    During $hashTable$ construction,  the continuous IDs of the neighbors in the largest outdegree vertex will experience minimum collision in $neighborList$.
    For reordering, the vertices need to be sorted by collective degree, and each edge needs to be scanned once. So the time complexity is $\mathcal{O}(|V|log|V|+|E|)$.

\end{itemize}

\begin{table*}[htbp]
    \setlength{\tabcolsep}{3.5mm}
    \centering
    \begin{tabular}{lrrrrrrrrrrrrr}
    \hline
          & \multicolumn{1}{c}{RM} & \multicolumn{1}{c}{RA} & \multicolumn{1}{c}{3D} & \multicolumn{1}{c}{MA} & \multicolumn{1}{c}{CP} & \multicolumn{1}{c}{OR} & \multicolumn{1}{c}{WK} & \multicolumn{1}{c}{FS} & \multicolumn{1}{c}{TW} & \multicolumn{1}{c}{IT} & \multicolumn{1}{c}{GH} & \multicolumn{1}{c}{CW} & \multicolumn{1}{c}{UK} \\\hline
		  BS    & 10    & 8     & 5     & 12    & 9     & 29    & 50    & 42    & 146   & 114   & 335   & 156   & 358 \\
		  CO    & 10    & 8     & 5     & 12    & 9     & 15    & 14    & 17    & 15    & 14    & 18    & 16    & 23\\
		  CO + RO (IN) & 10    & 8     & 4     & 9     & 8     & 15    & 15    & 16    & 17    & 13    & 16  & 15  &  22\\
		  CO + RO (OUT) & 9     & 8     & 4     & 9     & 6     & 14    & 15    & 16    & 15    & 12    & 15 & 15 & 20\\
		  CO + RO (OUT) + PA & - & - & - & - & - & - & - & - & - & - & - & 11 & 13\\\hline    
	\end{tabular}%

  \caption{Max collision, where BS, CO, RO (IN), RO (OUT), and PA stand for baseline, co-optimization, reordering, indegree based reordering, outdegree based reordering, and partition, respectively.}
    \label{tab: collision}%
  \vspace{-0.1in}
 \end{table*}%

\vspace{-0.1in}
\subsection{Virtual Combination for Workload Balancing}\label{subsec:VB}


\begin{figure}[h]
\vspace{-0.1in}
    \centering
    \includegraphics[width=\linewidth]{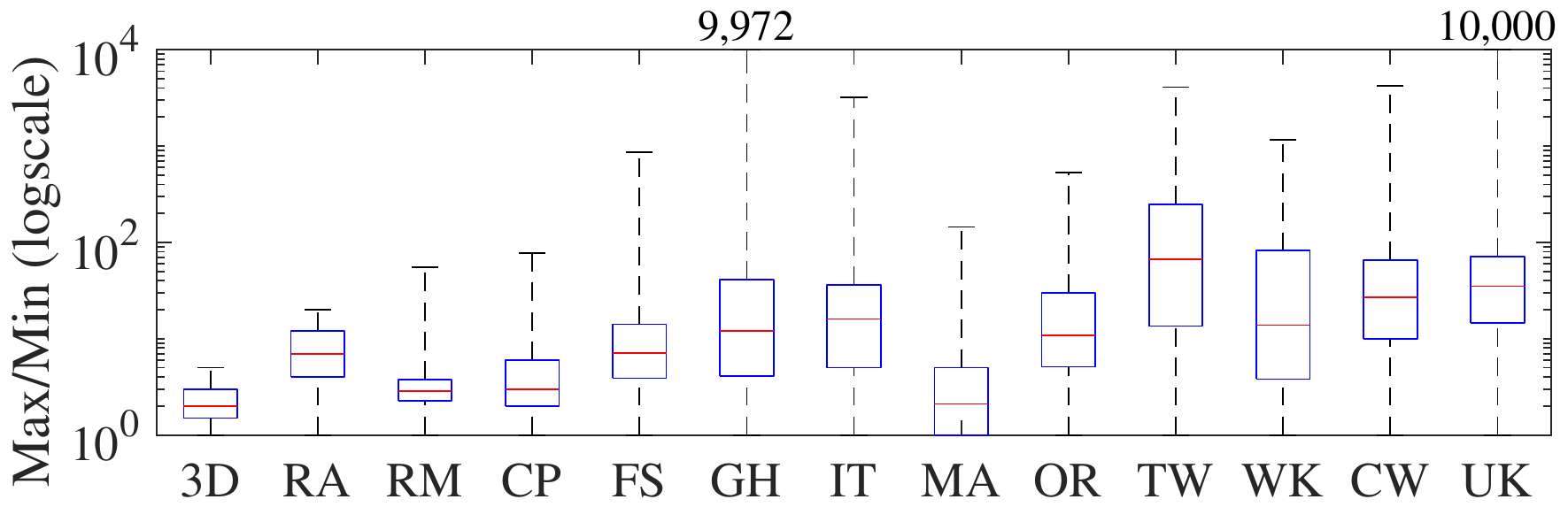}
    \vspace{-.015in}
    \caption{\final{The intra-vertex workload imbalance.} 
    \vspace{-0.05in}
    }
	\label{fig:boxplot degree}
\end{figure}

Intra-vertex workload imbalance hampers the performance of vertex-centric hashing. Figure~\ref{fig:boxplot degree} shows the workload imbalance with a boxplot of the ratio of maximum degree ($v$)/minimum degree ($v$) for each $u$, where $v\in u.neighborList$. {Particularly, the average of all the medians is {16} across graphs, with the average of the maximum as {2,648}}. We consequently need to accommodate each $v\in u.neighborList$ distinctly. 

There mainly exist two conventional resolutions to solve such an intra-vertex workload imbalance problem.
i) \textbf{Warp-centric} uses a warp of threads to work on one vertex so that the workload imbalance issue can be mitigated across all threads in a warp. However, this approach would suffer from thread under utilization since the average of the median is 16 which is smaller than the size of a warp. 
ii) \textbf{Subwarp}~\cite{polak2016counting,hong2011accelerating} is a straightforward optimization to mitigate the idling thread problem in the warp-centric approach. 
Basically, this method divides a warp into several subwarps and assigns each subwarp to one $neighborList$. This technique reduces the number of idling threads, but not entirely. Furthermore, some neighbors might have $neighborList$ whose sizes are larger than the subwarp sizes, leading to yet another workload imbalance concern.

{\hash} aims to ultimately resolve the workload imbalance and thread idling issues. We introduce two possible designs, i.e., \textbf{physical} and \textbf{virtual combinations}. The former one copies all the $neighborList$ of $v$ into a single combined array and processes them together. 
However, copying all the $neighborList$s of $v$ into one array could consume both nontrivial time and memory~\cite{liu2019simd}.
\textbf{Virtual combination} avoids copying the $neighborList$ of $v$ into a combined array via \textit{on-the-fly calculation of the 2-hop neighbor indices for each thread}. Particularly, assuming we are working on vertex $u$, because thread $i$ copies the 2-hop neighbors of $u$ to indices $i$, $i+32$, etc., in the combined array, we simply need to find which $v\in neighborList(u)$ contains neighbors that will be copied to those indices. Once $v$ is identified, we can further calculate which neighbor of $v$ will be copied by thread $i$. This way, we find the neighbors for thread $i$.

\begin{figure}[h]
	\centering
    \includegraphics[width=.9\linewidth]{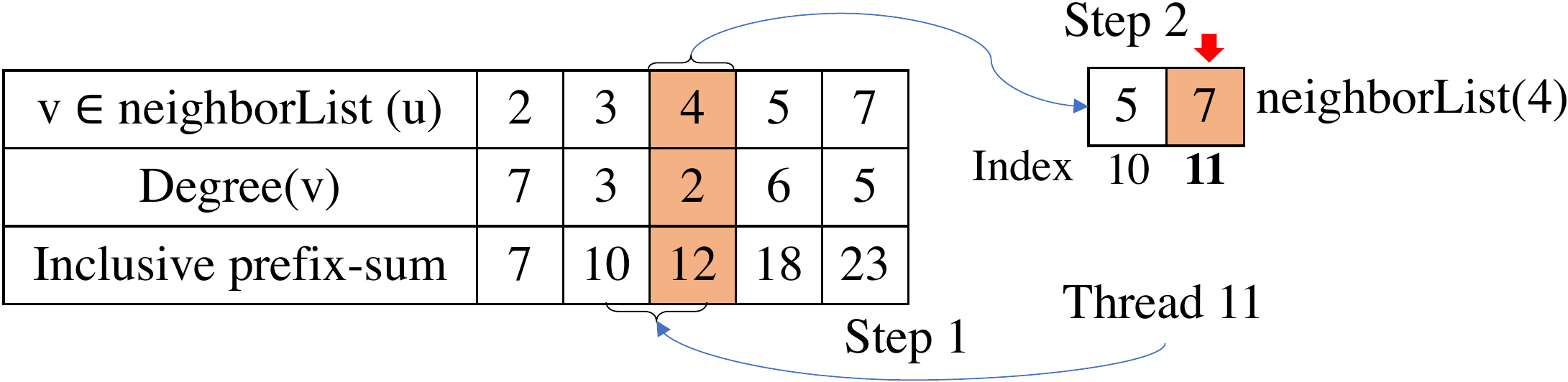}
	\caption{Virtual combination.\vspace{-.1in}}
	\label{fig:combination}
\end{figure}

Figure~\ref{fig:combination} uses an example to aid the understanding. Assuming vertex $u$ has neighbors \{2, 3, 4, 5, 7\} and their degrees are \{7, 3, 2, 6, 5\}, leading to the inclusive prefix-sum of these degrees as \{7, 10, 12, 18, 23\}. For thread 11, its index of interest is 11. At step 1, this thread finds $v=4$ which contains the neighbors that this thread will process because $v=4$'s neighbor range is [10, 12) in the combined array. 
At step 2, this thread computes that the second neighbor of vertex $v=4$ becomes the neighbor stored at index 11 in the combined array. Thus, the second neighbor, i.e., 7, will be processed by thread 11.

\vspace{-0.1in}
\subsection{Collision and Workload Imbalance Co-optimization}\label{sec:Co-opt}

\begin{figure}[ht]
	\vspace{-0.1in}
    \centering
    \includegraphics[width=\linewidth]{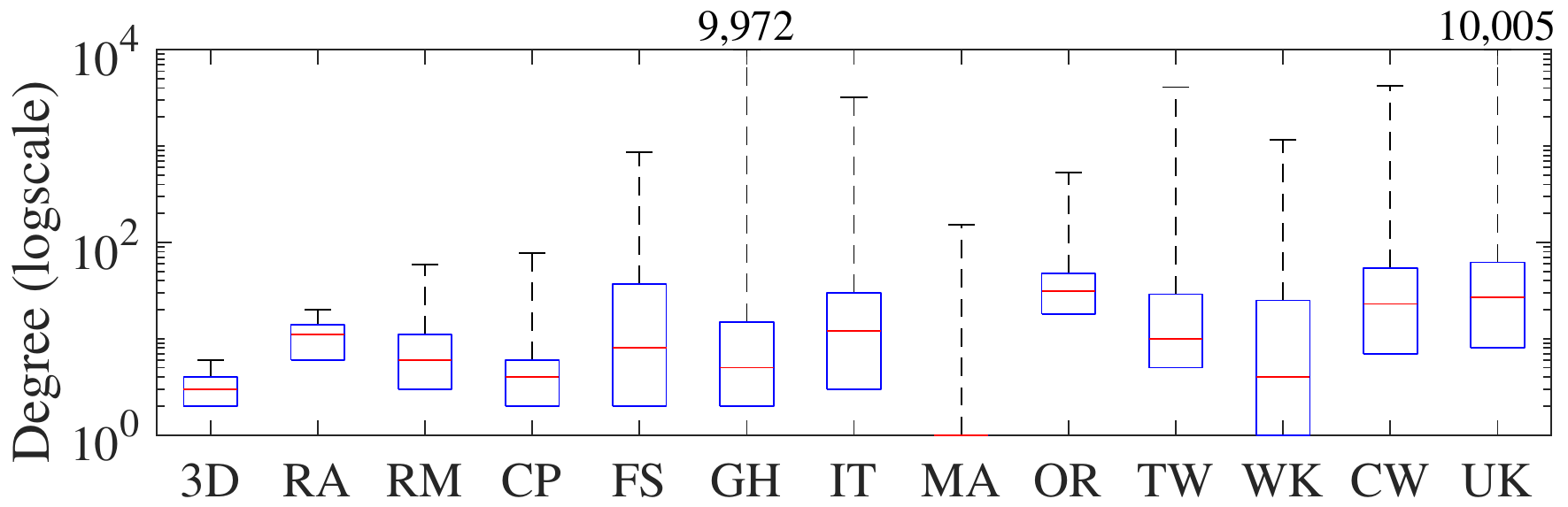}
    \vspace{-.05in}
    \caption{\final{Degree distribution of each graph after orientation.} }
    \label{fig:boxplot_udegree}
    \vspace{-.05in}
\end{figure}
    
This part is motivated by the key observation in Figure~\ref{fig:boxplot_udegree}, that is, even after orientation~\cite{shun2015multicore}, various vertices present different degrees. Particularly, the difference of maximum and minimum degrees can reach as high as {10,005} for {GH} graph.
These degree differences manifest as differences in $hashTable$ construction cost, collision, and workload. 

We advocate assigning different computing and shared memory resources for vertices with dissimilar degrees. Particularly, we assign a CTA with more shared memory for larger degree vertices, a warp and a smaller amount of shared memory for smaller degree vertices. Based upon our evaluation, we label vertices with $degree > 100$ as a large vertex for better performance.
Note that we do not need to process vertices with $degree<2$ since a vertex needs at least two neighbors to enumerate a triangle. It is important to note that large degree vertices obtain not only more threads to construct \textit{hashTable} and conduct intersection, but also more shared memory to cache hash buckets.

Table~\ref{tab: collision} studies the maximum collision changes with respect to various optimizations. Particularly, the maximum collision in the baseline version is larger than our threshold (128) in TW (146), and GH (335) graphs. However, after our collision-reducing optimizations, the maximum collision is no more than 16. 
For the extremely large graphs (i.e., CW and UK) whose sizes are bigger than GPU memory, the partitioning scheme can reduce the maximum collision below 16, in addition to the help from CO and RO optimizations.

\begin{figure}[h]
    \vspace{-0.1in}
    \centering
    \includegraphics[width=\linewidth]{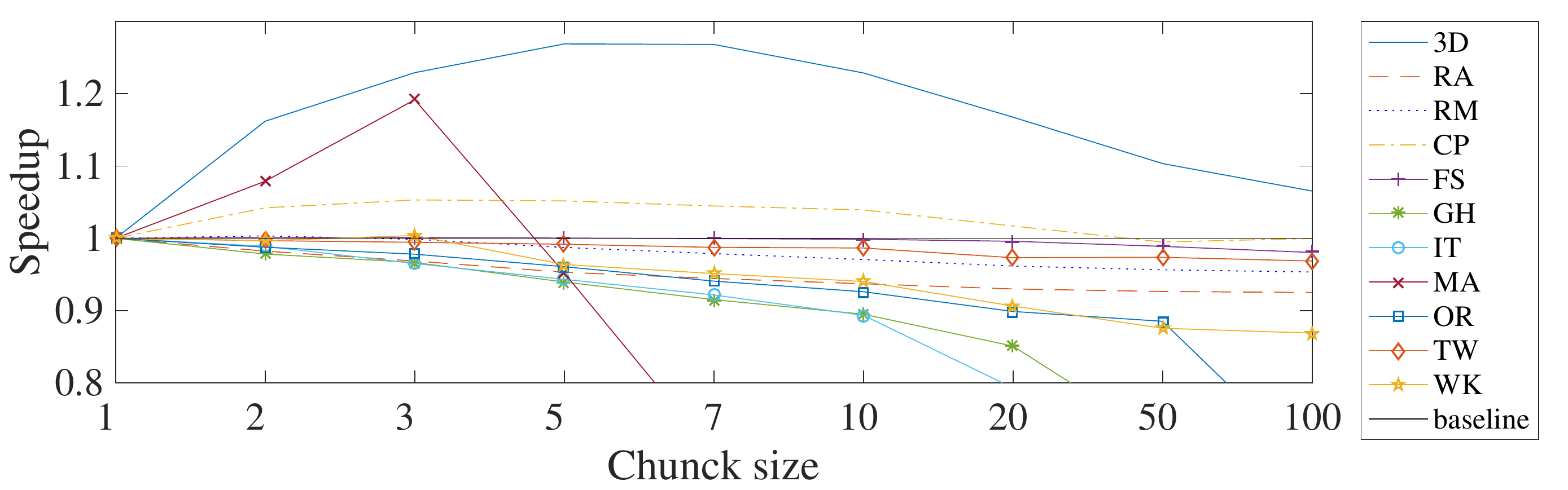}
    \vspace{-.1in}
    \caption{Chunk size selection.}    
    \vspace{-.1in}
    \label{fig:chuncksize}
\end{figure}

While degree-aware resource assignment can mitigate the workload imbalance, there still exists inter-vertex workload imbalance. 
We further introduce an atomic operation-based dynamic workload assignment to balance the workload. In this design, each warp/CTA gets a chunk of vertices atomically at a time. Depending upon the graph, the chunk size can be dissimilar. Figure~\ref{fig:chuncksize} shows the performance impacts of various chunk sizes. For sparse graphs, e.g., 3D, CP, and MA, a larger chunk size leads to 27\% (3D, chunk size = 7), 5\% (CP, chunck size =3) and 19\% (MA, chunk size = 3) speedup. 
For the rest of the graph datasets, chunk size = 1 gives the best performance.

\begin{figure*}[t]
	\centering
     \vspace{-0.1in}
    \includegraphics[width=.9\textwidth]{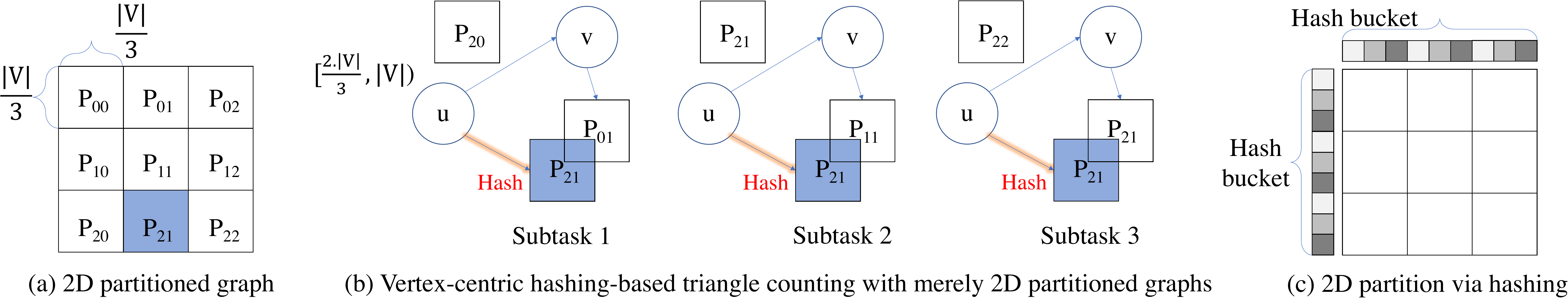}
    \caption{2D partition for vertex-centric triangle counting as well as hashing-based 2D partition.
    }
    \label{fig:Graph partition}
    \vspace{-.1in}
\end{figure*}

\section{Scalable Triangle Counting via Graph and Workload Collaborative Partitioning}
\label{sec:scale}

This section tackles the scalability challenge for triangle counting via a graph and workload collaborative partitioning design. As shown in Table~\ref{tab:task number}, with $n^2$ graph partitions, as long as each of them can fit in GPU memory, {\hash} can scale up to $m\cdot n^3$ GPUs, where $m$ and $n$ are the numbers of workload and graph partitions, respectively.

\begin{table}[h]
    \renewcommand\arraystretch{1.1}
    \centering
    \scalebox{1}
    {
        \begin{tabular}{|c|c|c|c|c|}
            \hline
                Partition approach & Workload  & Graph & Workload\&Graph \\\hline
            \#Tasks (i.e. GPUs) & $m$  & $n^3$ &$m\cdot n^3$\\\hline
            \#Graph partitions & 1  & $n^2$ &$n^2$\\\hline
            \multirow{2}{*}{Average \#edges/task} & \multirow{2}{*}{$|E|$}     & \multirow{2}{*}{$\frac{3*|E|}{n^2}$}& \multirow{2}{*}{$\frac{3*|E|}{n^2}$}\\
            & & &
            \\\hline
        \end{tabular}%
    }
    \vspace{.1in}
    \caption{Graph and workload collaborative partition vs. traditional workload alone, and graph partition alone methods, where $m$ and $n$ are numbers of workload and graph partitions, respectively. 
    \vspace{-0.1in}
    }
    \label{tab:task number}%
\end{table}%

\vspace{-0.1in}
\subsection{Workload Partitioning}
\label{subsec:workload}


Workload Partitioning assumes the entire CSR format of the graph can fit in a GPU memory so that we can directly duplicate the entire graph across all the GPUs. Subsequently, we only need to focus on the workload distribution across GPUs.
An intuition of workload partition is to distribute all the vertices into $m$ subsets. Afterward, each GPU can work on one such subset and count the triangles. 


{\hash} achieves balanced workload assignment through hashing on a slight modification to our aforementioned graph reordering techniques (Section~\ref{subsec:reorder}). Particularly, instead of assigning continuous IDs to all $v\in u.neighborList$ in the prior design, we first divide $v$ into three subsets: [0, 2), [2, 100] and (100, $+\infty$). Subsequently, we assign continuous IDs to $v\in$ (100, $+\infty$) from $u$ first, then $v\in$ [2, 100], and finally $v\in$ [0, 2). One can exploit radix hashing to distribute the vertices to various GPUs evenly. {For instance, assuming there are {$g$} GPUs, for GPU i, we let it process vertex $u$ such that {$u\%g=i$}.} Because our reordering approach assigns continuous IDs to vertices with similar degrees, radix hashing ensures that vertices of similar workloads are evenly disseminated to across GPUs. 
Note, this design is distinct from the traditional 1D/2D partitioning efforts~\cite{hu2017trix,hu2018tricore,jia2017distributed} that assign a continuous range of vertices to each GPU. {And, in this case, the collective reordering becomes an collective-degree-then-outdegree based reordering}.

\vspace{-0.1in}
\subsection{Graph Partition}

Chances are the entire CSR of a graph might not fit in the GPU memory, e.g., UK graph~\cite{BMSB} evaluated in this paper consumes more than {160 GB} memory. When this happens, we need to partition the graphs into smaller subgraphs so that each of them can fit in GPU memory. This also underscores the weakness of prior projects~\cite{hu2017trix,hu2018tricore} that need both edge list and $neighborList$ for triangle counting.
To better illustrate the design, we first review what information is needed in vertex-centric hashing-based triangle counting on a single GPU. Particularly, we need three $neighborList$: 

\begin{enumerate}[label=(\roman*)]
    \item $u$'s 1-hop $neighborList$ to construct $hashTable$.
    \item $u$'s 2-hop $neighborList$.
    \item $u$'s 1-hop $neighborList$ as sources to fetch the 2-hop $neighborList$ of bullet (ii).
\end{enumerate}

It is important to note that vertex-centric triangle counting focuses on the range of vertices. 
Particularly, for a vertex $u$ falling in a specific range, we can use all the partitions of that row to construct the $hashTable$, as well as the 1-hop neighbor to fetch the 2-hop neighbors. As shown in Figure~\ref{fig:Graph partition}, using u$\in[\frac{2\cdot|V|}{3}, |V|)$ as an example, we can use $P_{20}$, $P_{21}$ and $P_{22}$ to build the $hashTable$ and fetch 2-hop neighbors. However, using $P_{20}$, $P_{21}$ and $P_{22}$ together to fetch the 2-hop neighbor would result in fetching the entire graph.

\textbf{Partitions for 2-hop neighbors.} The good news is -- in order to extract a triangle, we only need the vertex range of the $hashTable$ to overlap that of the 2-hop neighbor. This helps reduce the number of fetched 2-hop neighbor partitions tremendously. For instance, for $P_{21}$ that is used for $hashTable$ construction, only 2-hop neighbor partitions whose destination vertices fall in $[\frac{|V|}{3}, \frac{2\cdot|V|}{3})$ are needed. In this example, only $P_{01}$, $P_{11}$ and $P_{21}$ are needed for 2-hop neighbors. Similarly for $P_{20}$ and $P_{22}$.

\textbf{Partitions for 1-hop neighbors.}
We further need to fetch the 1-hop neighbor partitions that are used to index the 2-hop $neighborList$. The key is that \textit{$u$'s 1-hop neighbors used to construct the $hashTable$ and the 1-hop neighbors used to index the 2-hop neighbors can be different.} If we force them to be the same, we will end up only intersecting the $hashTable$ with the diagonal partitions. Using $u$'s range of $[\frac{2\cdot|V|}{3}, |V|)$ as an example, the second partition, that is, $P_{21}$ from Figure~\ref{fig:Graph partition}(b) is used to construct the $hashTable$. We can use any partitions whose source vertices are in the range of $[\frac{2\cdot|V|}{3}, |V|)$ as the sources to index the 2-hop neighbors. In this case, $P_{20}$, $P_{21}$ and $P_{22}$ are the qualified partitions to index the 2-hop $neighborList$s.

In addition to soundness, this design is also complete because we exhaust all the possible 2-hop neighbor partitions for each $hashTable$ partition. As shown in Figure~\ref{fig:Graph partition}(b), for the 1-hop neighbor partition $P_{21}$ that is used to construct the $hashTable$ of the vertices under processing $u\in[\frac{2\cdot|V|}{3}, |V|)$, we use all the possible 1-hop neighbors, that is, $P_{20}$, $P_{21}$ and $P_{22}$, as sources to index the 2-hop neighbors. 

\vspace{-0.1in}
\subsection{Workload and Graph Collaborative Partition}\label{subsec:colla}

This section further integrates our graph partitioning technique with our workload partitioning design. Particularly, for the same $hashTable$, we distribute each 2-hop neighbor partition to one GPU, so that all GPUs work on different workloads of the same $hashTable$. We distribute three 2-hop neighbor partitions $P_{20}$, $P_{21}$ and $P_{22}$ - of the $hashTable$ partition $P_{21}$ across three GPUs. 
With total $n^3$ subtasks, we further divide each subtask into $m$ workload partitions in order to scale to $m\cdot n^3$ GPUs.

It is worthy of mentioning that, instead of using vertex range-based 2D graph partitioning as shown on the left side of Figure~\ref{fig:Graph partition}(a), {\hash} exploits hashing to generate the partitions in Figure~\ref{fig:Graph partition}(c).
For partition $P_{ij}$, it contains the edges $(u,v)$ where $u\%n=i$ and $v\%n=j$. 
As shown in Figure~\ref{fig:Graph partition}(c), we first exploit hashing to decide which row partition $u$ belongs to, subsequently, another hashing towards $v\in u.neighborList$ to decide which column partition each $v$ belongs to. Thanks to our reordering, our hashing-based partition warrants a roughly similar number of vertices and edges for each partition.

However, since using hashing to partition vertex set will lead to noncontinuous IDs for each partition that is detrimental to $hashTable$ construction, we reassign IDs by $\text{newID}=\lfloor\text{oldID}/n\rfloor$. In this way, the vertices IDs in each partition become continuous. And partitioning only needs to scan each edge once, resulting in a time complexity of $\mathcal{O}(|E|)$.

\textbf{Integrating partitioning with aforementioned optimizations.} \update{Here, the aforementioned optimizations are ``reordering'', and ``collision and workload imbalance co-optimization''. First, ``reordering'' is performed to ensure that the vertices in the same subset have continuous IDs.
Then, hashing-based partitioning can evenly partition the vertices in each subset.}
Second, ``co-optimization'' performed after partitioning in Section~\ref{sec:Co-opt} divides the vertices of each partition into three subsets by their degrees. The subsets (i.e., large vertices, small vertices, and omissible vertices) represent their workload. Since partitioning divides the $neighborList$ of each vertex, a vertex originally belonging to large vertices subset might change to the small vertices subset. 
To track which subset a vertex belongs to after partitioning, we propose a mapping between the partitioning and co-optimization steps. 

\begin{figure}[h]
    \vspace{0.1in}
    \centering
    \includegraphics[width=.9\linewidth]{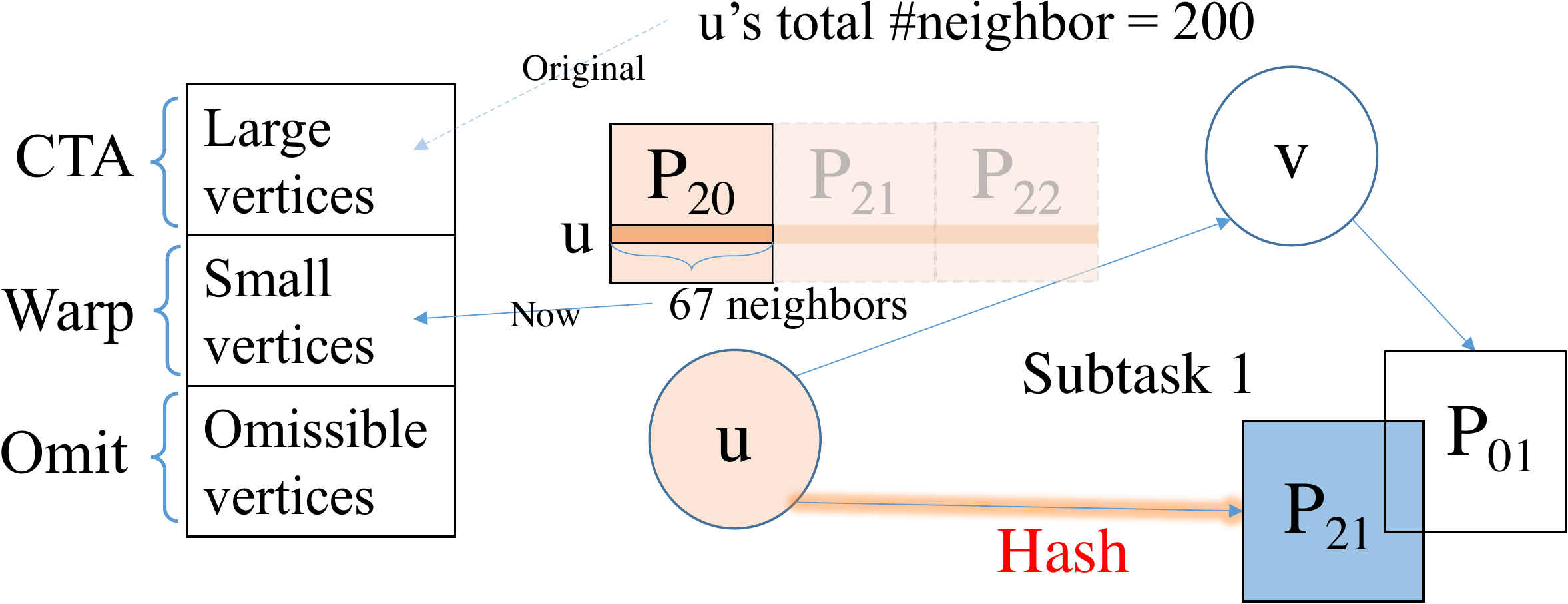}
    
    \caption{Integrating graph partitioning with collision and workload balance co-optimization in Section~\ref{sec:Co-opt}.
    }
    \label{fig:mapping}
\end{figure}

Figure~\ref{fig:mapping} uses the subtask 1 from Figure~\ref{fig:Graph partition}(b) to explain the idea. The $neighborList$ of vertex $u$ is divided into three partitions (i.e., $P_{20}$, $P_{21}$ and $P_{22}$) as show in Figure~\ref{fig:mapping}. For this specific subtask vertex $u$, $P_{20}$ determines the workload of vertex $u$. 
We assume $u$, in total, has 200 neighbors and belongs to the large vertices subset before partition. Because partitioning distributes $u$'s $neighborList$ across $P_{20}$, $P_{21}$ and $P_{22}$, we assume $u$ has 67 neighbors in $P_{20}$. In this case, $u$ should belong to the small vertices subset after partition.
Therefore, for subtask 1 in Figure~\ref{fig:Graph partition}(b), we treat $u$ as small vertex during co-optimization step.

\vspace{-0.05in}
\section{Evaluation}
\label{sec:eval}
{\hash}\footnote{Available at https://github.com/wzbxpy/TRUST} is implemented with around 1,500 lines of C++/CUDA code and compiled with CUDA Toolkit 10.2, g++ 7.4.0, MPICH-3.3, and the optimization flag is set to -O3. We evaluate {\hash} on two servers: i) a server with Intel(R) Xeon(R) Gold 6248 CPU with 40 cores, 512 GB main memory, and 8 V100 GPUs, each with 32 GB memory; ii) Summit supercomputer~\cite{ornl_summit} with 512 GB memory, powered by dual-socket 22-core POWER 9 processor along with 6 V100 GPUs, each of which installs 16 GB GPU memory. We use Summit only when evaluating the scalability of medium and extremely large graphs in Section~\ref{subsec:scalability}. For the remaining experiments, we use server (i). The runtime of triangle counting is measured once the graph is loaded on GPUs for comparison with state-of-the-art systems.


For MPI-based implementation with multi-GPUs, we use the maximum kernel time across all participating GPUs as the triangle counting time. {Unless} otherwise specified, the reported statistics are the average of ten runs. 

\vspace{-0.1in}
\subsection{{\hash} vs. State-of-the-Art}
  
This section compares {\hash} with two state-of-the-art triangle counting systems, i.e., Ligra~\cite{shun2013ligra,shun2015multicore} and TriCore~\cite{hu2018tricore}. Particularly, Ligra is a lightweight graph processing framework. We compile the Ligra source code with Intel CILK library to achieve peak performance and test Ligra on Intel(R) Xeon(R) Gold 6248 CPU with 40 cores and 512 GB main memory. TriCore is regarded as the optimal GPU-based triangle counting system that won the 2018 GraphChallenge champion~\cite{hu2018high}. TriCore and {\hash} run on a single V100 GPU. {Comparing the prices, one Intel(R) Xeon(R) Gold 6248 CPU costs around \$6,600~\cite{CPUprice} while a single V100 GPU costs around \$11,500~\cite{GPUprice}}.
  
\begin{figure}[h]
    \vspace{-0.1in}
    \centering
    \includegraphics[width=\linewidth]{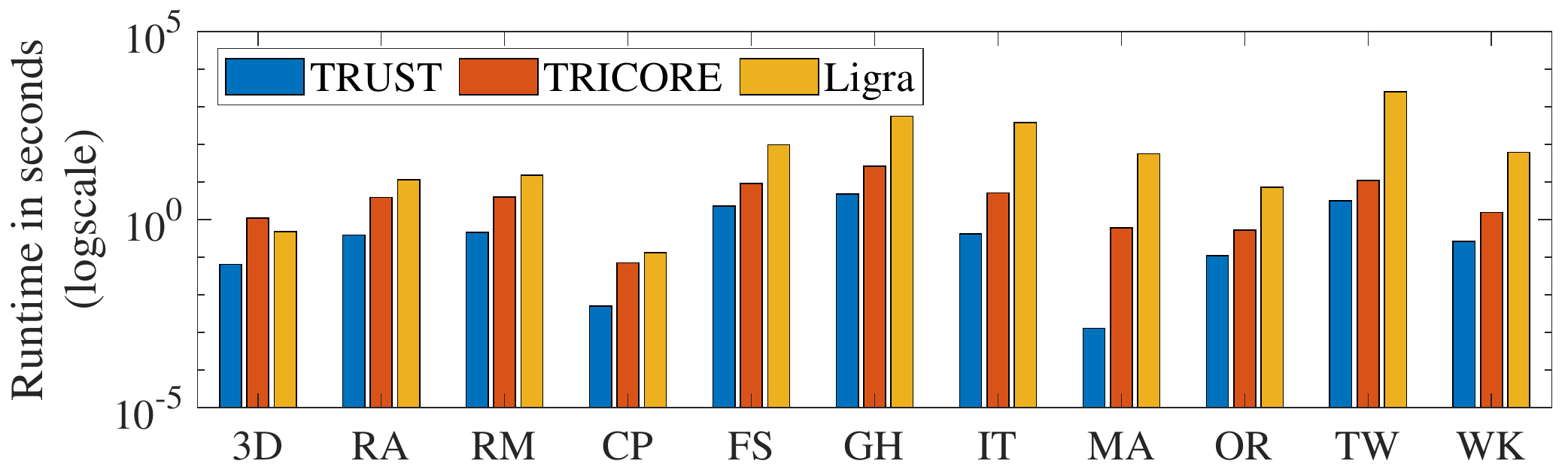}
    \caption{The runtime of {\hash}, TriCore, and Ligra. 
    \vspace{-0.05in}
    }
    \label{fig:compare_with_state_of_the_art}
\end{figure}

As shown in Figure~\ref{fig:compare_with_state_of_the_art}, {\hash} achieves 50.1$\times$ and 4,177.4$\times$ speedup on average over TriCore and Ligra, respectively. Comparing to TriCore, {\hash} achieves 465.0$\times$ speedup on the MA graph. For the remaining graphs, the speedup ranges from 3.4$\times$ (TW) to 17.2$\times$ (3D).
TriCore enjoys irregular graphs (like power-law graphs) but suffers from regular graphs like 3D, RA, and RM. The reason lies in the fact that TriCore is designed upon binary-search, which is more efficient when the degree differences between vertices are larger. 
Comparing to Ligra, {\hash} {beats} Ligra by 43,697.3$\times$ on the MA graph. For the remaining graphs, the speedup ranges from 7.4$\times$ (3D) to 919.7$\times$ (IT). The general trend is that {\hash} has significant margins over Ligra when the graph is larger and more irregular.


\vspace{-0.11in}
\subsection{{\hash} vs. GraphChallenge Champions}
\label{subsec:vs champion}

{This section compares {\hash} against H-INDEX~\cite{pandey2019h}, Bisson et al.~\cite{bisson2018update}, and Yacsar et al.~\cite{yacsar2018fast} which are the champions in 2018 and 2019 GraphChallenge~\cite{graphchallenge}. Yacsar et al. follow matrix-multiplication approach for triangle counting. Since Yacsar et al.~\cite{yacsar2019linear} is the updated and faster version of 2018 GraphChallenge champion~\cite{yacsar2018fast}, we only choose~\cite{yacsar2019linear} among~\cite{yacsar2018fast,yacsar2019linear} for comparison. H-INDEX proposes to hash the shorter $neighborList$ for triangle counting while Bisson et al. relies upon bitmap-based intersection to do triangle counting. Table~\ref{tab:compare with bisson} shows the speedup achieved by {\hash} over these three related works. Since Bisson et al. and Yacsar et al. have not open-sourced their source code, Table~\ref{tab:compare with bisson} only includes three large graphs for Bission et al. and two large graphs for Yacsar et al. that are presented in manuscripts~\cite{bisson2018update} and ~\cite{yacsar2019linear}, respectively. In the manuscripts, Bission et al. is evaluated on one V100 GPU, and Yacsar et al. is evaluated on a DGX machine equipped with eight V100 GPUs and CPU with 40 cores. Yacsar et al. also utilizes both CPU and GPUs to count the triangles.  
We run {\hash} and H-INDEX on one V100 GPU.  }
\begin{table}[htbp]
    \renewcommand\arraystretch{1.15}
    \centering
    \scalebox{0.7}{
    \begin{tabular}{|c|c|cc|cc|cc|}
        \hline
            & {\hash}  & Bisson et al. & Speedup & H-INDEX & Speedup& {Yacsar et al.} & Speedup \\ \hline
        FS    & 2.241s & 3.935s  & 1.76$\times$  & 12.001s  & 5.36$\times$ &{3.133s}& 1.39$\times$  \\
        {MA}    & 0.001s & 0.023s  & 18.44$\times$  & 0.044s  & 34.84$\times$ & -&- \\
        TW    & 3.158s & 3.626s  & 1.14$\times$  & 74.424s  & 23.57$\times$ & {4.582s}& 1.45$\times$ \\
        \hline
    \end{tabular}%
    }
        \vspace{0.05in}
    \caption{{\hash} vs. GraphChallenge champions. Note, Yacsar et al. uses eight V100 GPUs, while the rest of the projects use one V100 GPU.
    \vspace{-0.1in}
    }
    \label{tab:compare with bisson}%
    
\end{table}%

As shown in Table~\ref{tab:compare with bisson}, {\hash} constantly outperforms the champions. On average, {\hash} achieves 7.1$\times$ and 21.3$\times$ speedup over Bission et al. and H-INDEX, respectively. We also notice that the margin of {\hash} over Bisson et al. on TW and FS is relatively small as bitmap tends to work well for graphs with a relatively small number of vertices. 
Because of large bitmap sizes, Bisson et al. fails to handle the extremely large graphs (such as CW and UK), which are all supported by {\hash}. Comparing with Yacsar et al., {\hash}, even with $\frac{1}{8}$ of the GPUs, is 1.4$\times$ faster on average. \final{Further, comparing with DistTC~\cite{hoang2019disttc}, a recent distributed triangle counting on GPUs, DistTC with 16 P100 GPUs is slower (3.92s in TW and 2.49s in FS) than {\hash} with 1 V100 GPU. Note, since DistTC is not a GraphChallenge champion, we do not include this result in Table~\ref{tab:compare with bisson}}.

\vspace{-0.1in}
\subsection{Impact of Various Optimizations}
\label{subsec: impact of various opt}

\begin{figure}[h]
    \vspace{-0.2in}

    \centering
    \subfloat[Smaller impacts.]{
        \hspace{-.1in}\includegraphics[width=.61\linewidth]{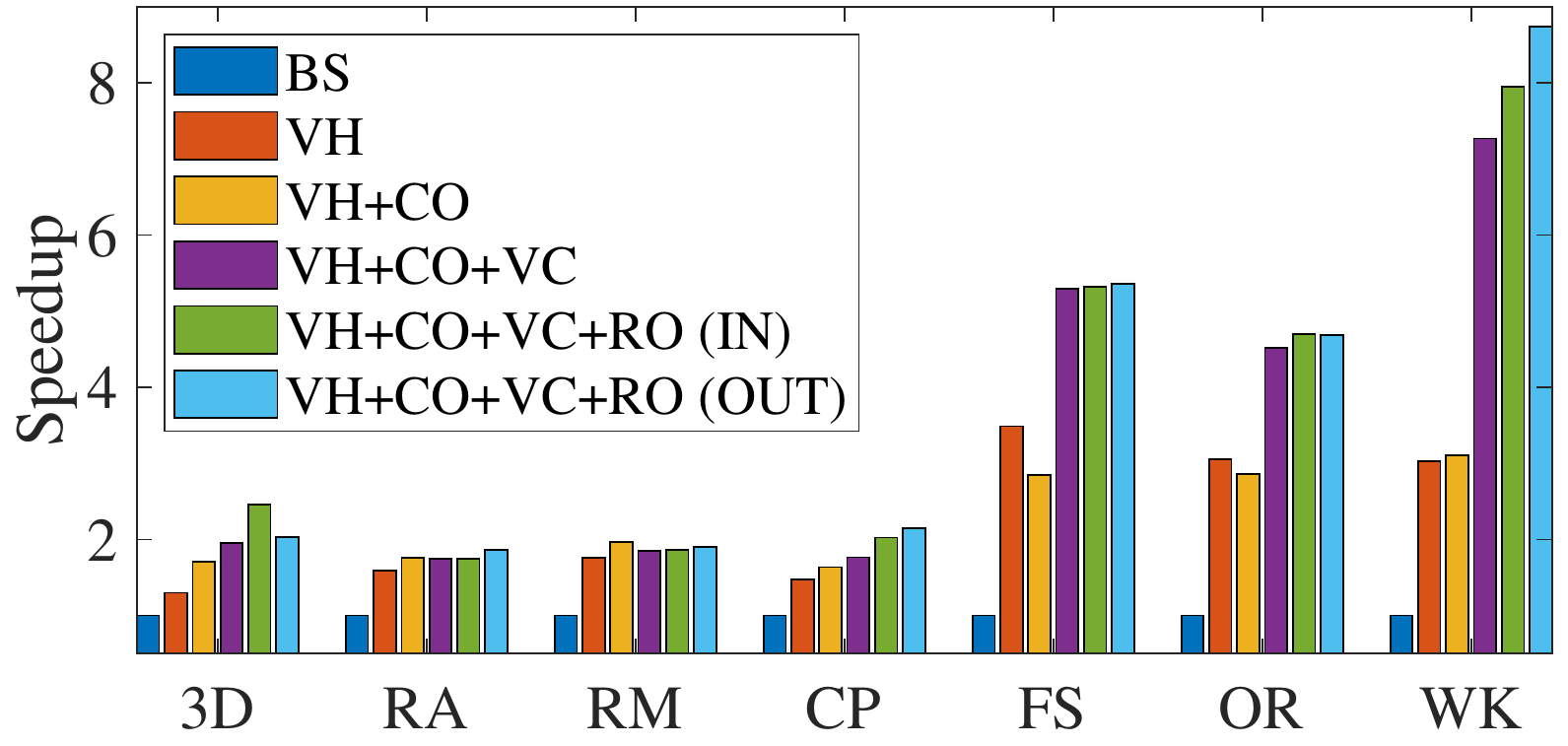}
    }
    \subfloat[Larger impacts.]{
        \includegraphics[width=.39\linewidth]{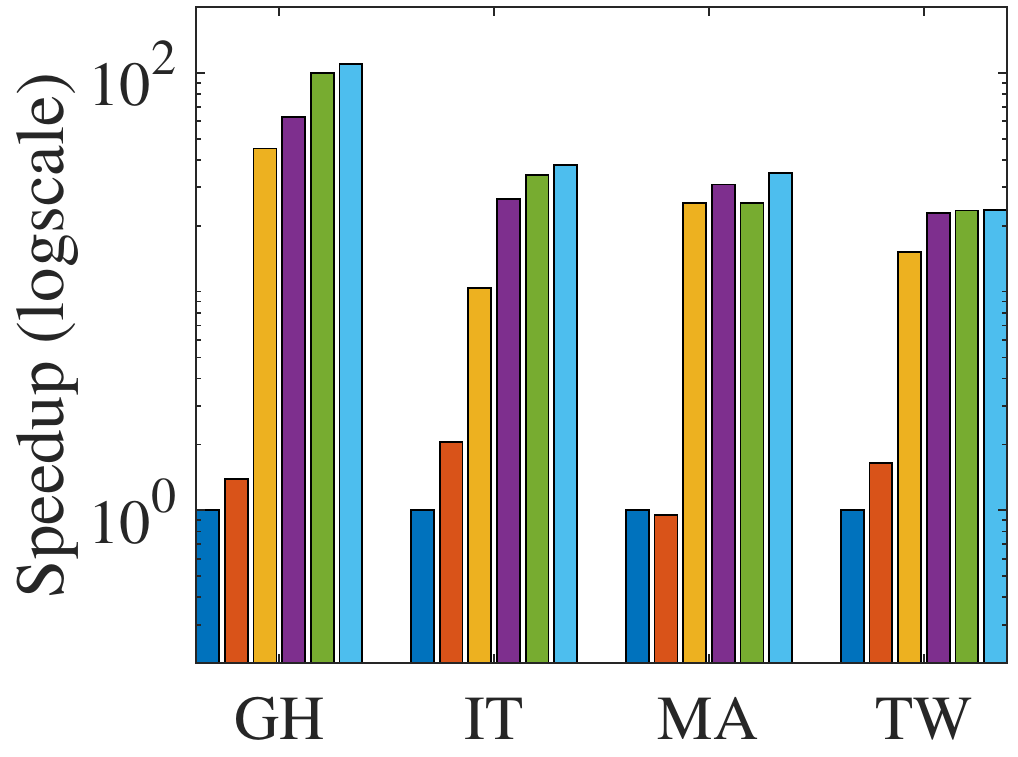}
    }
    \vspace{0.05in}
    \caption{Performance impacts of VH (vertex-centric hashing), CO (co-optimizing workload imbalance and hash collision), VC (virtual combination), and RO (vertex reordering). H-INDEX is used as the baseline (BS) for comparison.
    \vspace{-0.1in}
    }
    \label{fig:balance_opt}

\end{figure}

Figures~\ref{fig:balance_opt}(a) and~\ref{fig:balance_opt}(b) show the impacts of various optimizations categorized in terms of speedup. VH (vertex-centric hashing) achieves, on average, 2.0$\times$ speedup comparing with the baseline and upto 3.5$\times$ for FS graph. In contrast, for MA graph, VH is 5\% slower as most of the vertices in MA graph have $degree<2$. The reason is that the overheads of handling workload imbalance in vertex-centric design outweigh the benefit of $hashTable$ construction. 
CO (co-optimizing workload imbalance and hash collision) achieves only 1\% speedup on small impact graphs but achieves 18.0$\times$ speedup on large impact graphs as it balances the workload of highly skewed graphs. But CO is slower on FS (18\%) and OR (6\%) graphs as using CTA to process vertex leads to more idling threads due to the small workload. With the addition of VC (Virtual Combination), we observe another 50\% speedup on average across all graphs. However, as the degree distribution of RM and RA graphs are suitable for warp-centric processing, VC affects the performance for those graphs slightly.
Furthermore, we test two RO (vertex reordering) methods: Indegree (IN) and Outdegree (OUT). IN and OUT achieves 11\% and 18\% speedup on average across  all graphs, respectively. In most of the graphs, OUT outperforms IN.

\vspace{-0.1in}
\subsection{Profiling Reordering and Workload Balancing}
\label{subsec: profile RO and VC}

\begin{figure}[ht]
	\vspace{-0.15in}
    \centering
    \includegraphics[width=\linewidth]{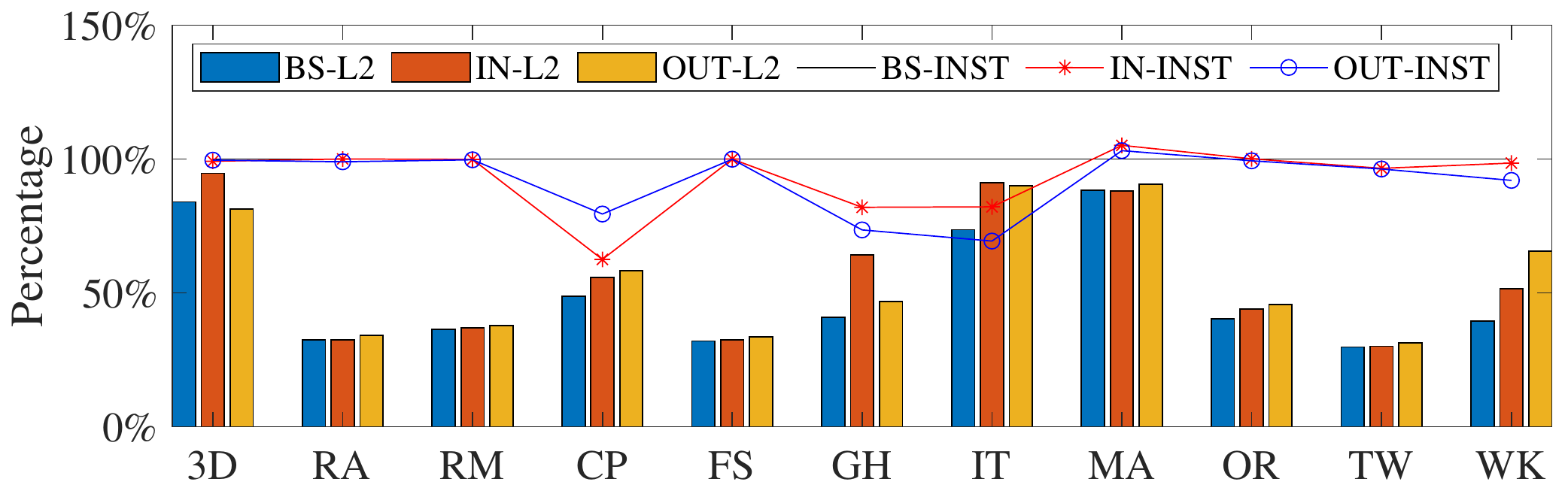}
    \caption{Percentage of L2 cache hit rate (L2) and warp level instructions for shared loads (INST) compared with baseline (BS) for IN and OUT reordering techniques.
    }
	\label{fig:reorder result}
	\vspace{-0.05in}
\end{figure}

\textbf{Profiling vertex reordering}. Figure~\ref{fig:reorder result} further profiles IN and OUT reordering techniques. Degree sorting technique~\cite{balaji2019combining} is used as the baseline (BS). We use Nvprof~\cite{nvprof} to profile {\hash}'s reordering techniques. 
The performance gain of reordering can be measured from two aspects: i) reduction of max collision in $hashTable$ and ii) improvement in the data locality of the $neighborList$. We profile max collision with warp level instructions for shared loads (INST), with the general idea being fewer collisions results in fewer memory reads. For profiling improvement in data locality, we use L2 cache hit rate (L2). Figure~\ref{fig:reorder result} shows that IN reduces INST by 6.8\% and improves L2 by 5.2\% on average. Similarly, OUT reduces INST by 8.1\%  and improves L2 by 6.3\% on average. 



\textbf{Profiling workload balancing}.
We perform another experiment to test {\hash}'s four intra-vertex workload balancing methods - warp-centric (WC), subwarp (SW), physical combination (PC), and virtual combination (VC). For SW, we test SW of size 8 and 16. 
\begin{figure}[h]
	\vspace{-0.1in}
    \centering
    \includegraphics[width=\linewidth]{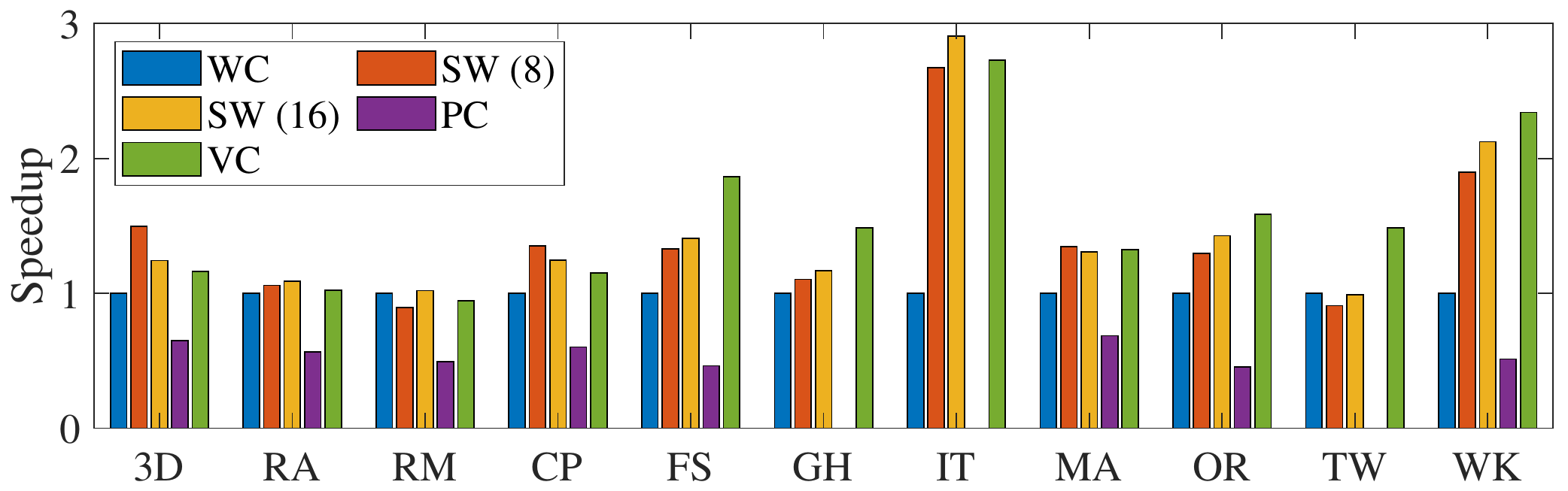}
    \caption{Profiling intra-vertex workload balancing methods.
    \vspace{-.1in}
    }
	\label{fig:Intra_balance_result}
\end{figure}

Figure~\ref{fig:Intra_balance_result} shows the speedup for different methods with WC as the baseline. On average, SW provides 40\% and 45\% speedup for subwarp of size 8 and 16, respectively. But we also observe that different graphs prefer dissimilar subwarp sizes, making it hard to pick one method for all graphs. When it comes to PC, it is 45\% worse than the baseline on average, blaming the cost of moving various $neighborList$s into a gigantic array.
VC achieves 55\% speedup over WC on average. Note that VC is slightly worse than SW (8) for relatively small degree graphs, such as 3D, and CP graphs and SW (16) for IT, RA, and RM graphs. However, SW (8) and SW (16) are significantly worse than VC for the rest of the graphs. Although the optimal SW sizes could yield the best performance, considering the difficulty with SW in selecting the correct SW size (8 or 16), {\hash} chooses the VC for intra-vertex workload balancing.

\vspace{-0.1in}
\subsection{{\hash} Scalability}
\label{subsec:scalability}
\vspace{-0.05in}

In this section, we discuss the scalability of {\hash} with the increase of GPUs.
For small and medium graphs, the number of workload partitions is equal to the number of available GPUs, i.e., $m= \#\text{GPUs}$. For extremely large graphs, each graph needs to be partitioned into smaller partitions which can fit in the GPU global memory. To achieve the best performance, 
we compute the smallest $n$ that satisfies $\frac{3*|E|}{n^2}*\text{edge size}<\text{GPU memory size}$. Then, we set $m=\#\text{GPUs}/n^3$. For the experiment, we set $n=8$, $m=1$ for 512 GPUs, and $n=8$, $m=2$ for 1,024 GPUs. 

\begin{figure}[t]
    \vspace{-0.2in}

    \centering
    \subfloat[Small graph scalability.]{
      \includegraphics[width=.45\linewidth]{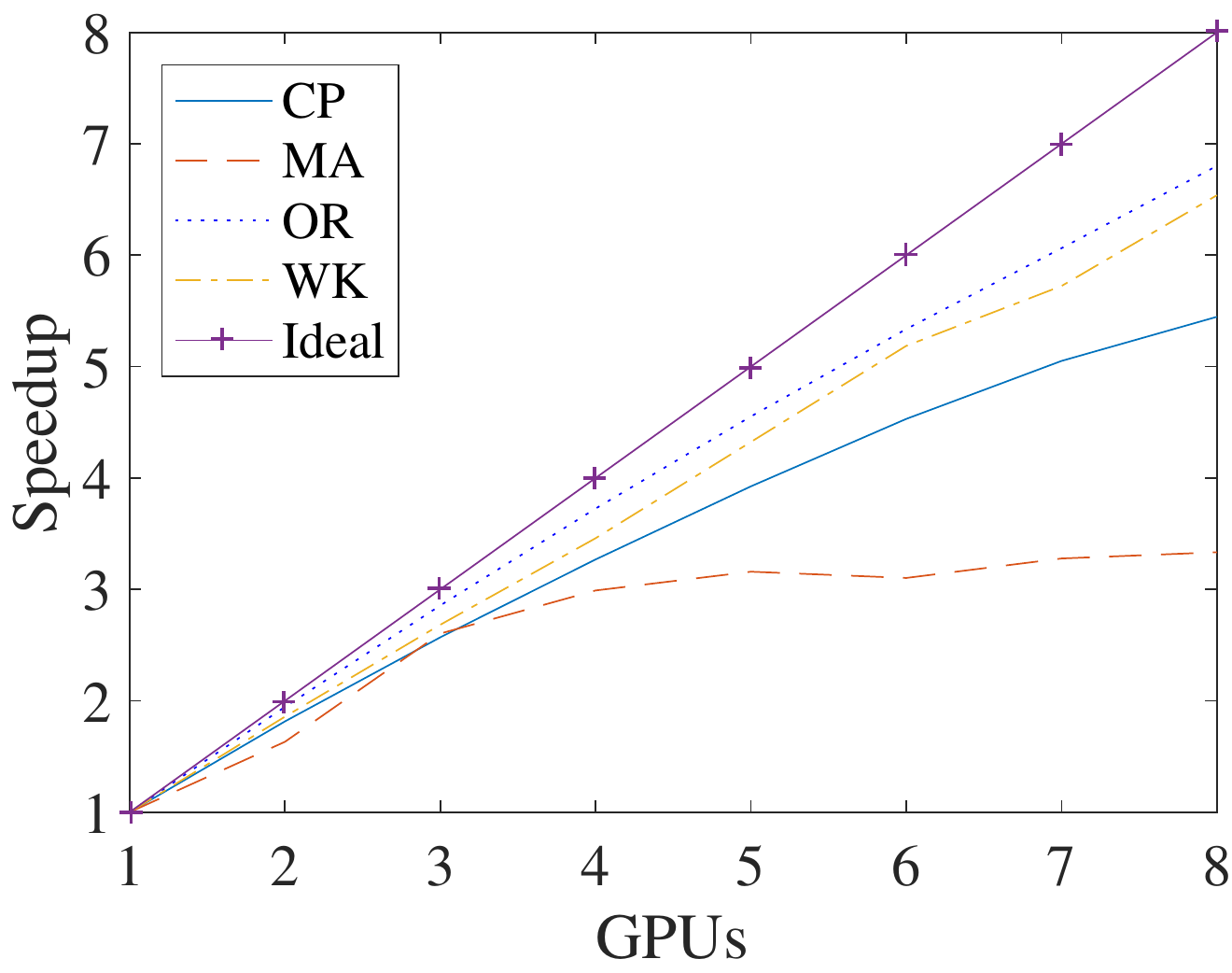}
    }
    \subfloat[Medium graph scalability.]{
        \includegraphics[width=.49\linewidth]{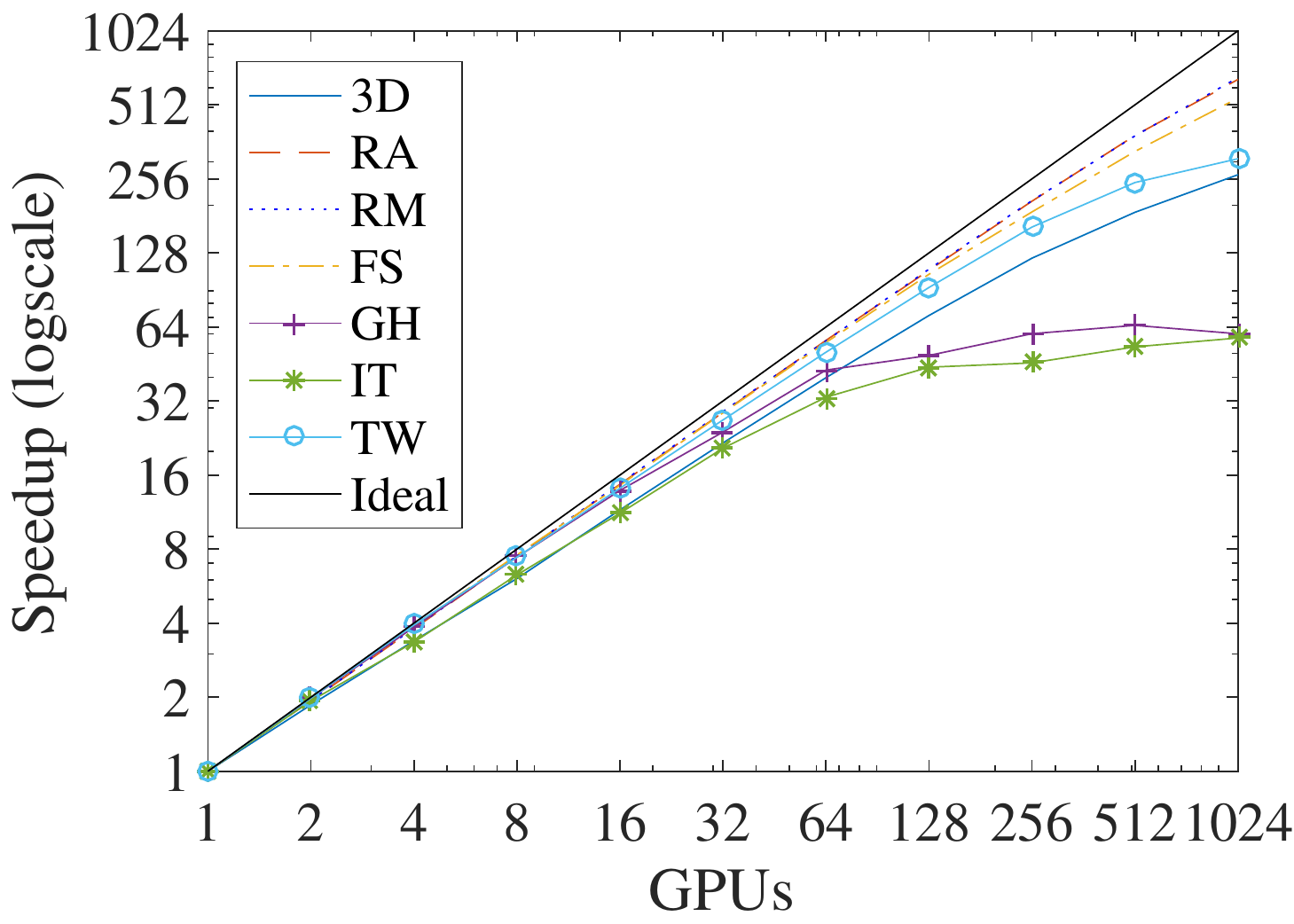}
    }
    \caption{Scalability for small and medium graphs.
    \vspace{-0.1in}
    }
    \label{fig:scalability}
\end{figure}

\textbf{Small {graphs}}. As shown in Figure~\ref{fig:scalability}(a), {\hash} achieves 3.3$\times$ to 6.8$\times$ speedup from 1 to 8 GPUs for four small graphs. For OR and WK graphs, {\hash} achieves almost linear scalability. 
In case of smaller workloads, like the MA graph, the scalability is limited when the computation resource is more than the number of tasks. 
In this situation, the runtime time is limited by the specific warp or CTA that processes the largest vertex.

\textbf{Medium {graphs}}. Figure~\ref{fig:scalability}(b) shows the scalability of $\hash$ for graphs of medium size. Particularly, {\hash} achieves 649.3$\times$ and 660.3$\times$ speedup for RA and RM graphs with 1,024 GPUs, respectively. For the rest of the graphs, the speedup is limited by their smaller workloads as discussed earlier.


\begin{table}[ht]
    \centering
    \renewcommand\arraystretch{1.2}
    \scalebox{0.85}{
    \begin{tabular}{|c|c|c|c|c|c|c|}
        \hline
        \#GPU& \multicolumn{2}{|c|}{512 GPUs} & \multicolumn{2}{|c|}{1,024 GPUs}&\multirow{2}{*}{Space IR} & LiteTe \\ \cline{1-5}  \cline{7-7}
        Measure & Time & Time IR &Time & Time IR& &Time IR\\ \hline
        CW &  0.15532s &1.10734& 0.08181s&1.10647& 1.06411 & - \\
        UK &  0.21023s &1.11559& 0.10942s&1.11919& 1.01359 & \update{1.70}\\ \hline
    \end{tabular}
    }
    \vspace{.1in}
    \caption{Scalability of extremely large graphs with 512 and 1,024 GPUs. Here, IR is short for imbalance ratio. Thus, Time IR = max time/min time. And Space IR = max partition size/min partition size.
    \vspace{-0.1in}
        }
    \label{tab:large graph partition}%
\end{table}

\textbf{Extremely large {graphs}}. As shown in Table~\ref{tab:large graph partition}, {\hash} achieves, on average, 1.9$\times$ speedup on CW and UK graphs while scaling from 512 to 1,024 GPUs. Further, looking into the imbalance ratio (IR), we observe that {\hash}'s graph partitioning achieves desirable workload (Time IR) and space (Sapce IR) balance. Particularly, both time IR and space IR lie between 1 and 1.1 for both graphs on 1,024 GPUs. In contrast, LiteTe~\cite{zhang2019litete}, which uses range-based partitioning scheme, has a much higher time IR, i.e., 1.7 for UK graph.

\section{Conclusion}
\label{sec:conclusion}
This paper introduces {\hash} that reloads triangle counting on GPUs. Particularly, it introduces vertex-centric hashing-based algorithm, collision and workload balancing optimizations, and workload and graph collaborative partitioning techniques. Taken together, {\hash}, to the best of our knowledge, is the first work that advances triangle counting beyond the trillion TEPS rate.

\section*{Acknowledgement}
We thank the anonymous reviewers for their helpful suggestions and feedback. This research is supported in part by the National Science Foundation CRII award No. 2000722, CAREER award No. 2046102 and  
the Exascale Computing Project (17-SC-20-SC), 
a collaborative effort of the U.S. Department of Energy Office of Science and the National Nuclear Security Administration. 
This research also is supported in part by the National Key R\&D Program of China 2018YFB1003505, the National Natural Science Foundation of China under Grant Numbers 61772265, 61802172, and 62072228. 
Sheng Zhong is supported in part by NSFC-61872176, The Leading-edge Technology Program of Jiangsu Natural Science Foundation (No. BK20202001), and National Key R\&D Program of China under Grant 2020YFB1005900. 
This research used resources of the Oak Ridge Leadership Computing Facility, which is a DOE Office of Science User Facility supported under Contract DE-AC05-00OR22725. This material is based upon work supported in part by the U.S. Department of Energy, Office of Science (DOE-SC) at Brookhaven National Laboratory, which is operated and managed for DOE-SC by Brookhaven Science Associates under contract No. DE-SC0012704.



	{
		\bibliographystyle{unsrt100.bst}
		\bibliography{references}
  }
\vspace{-0.5in}
\begin{IEEEbiography}[{\includegraphics[width=1in,height=1.25in,clip,keepaspectratio]{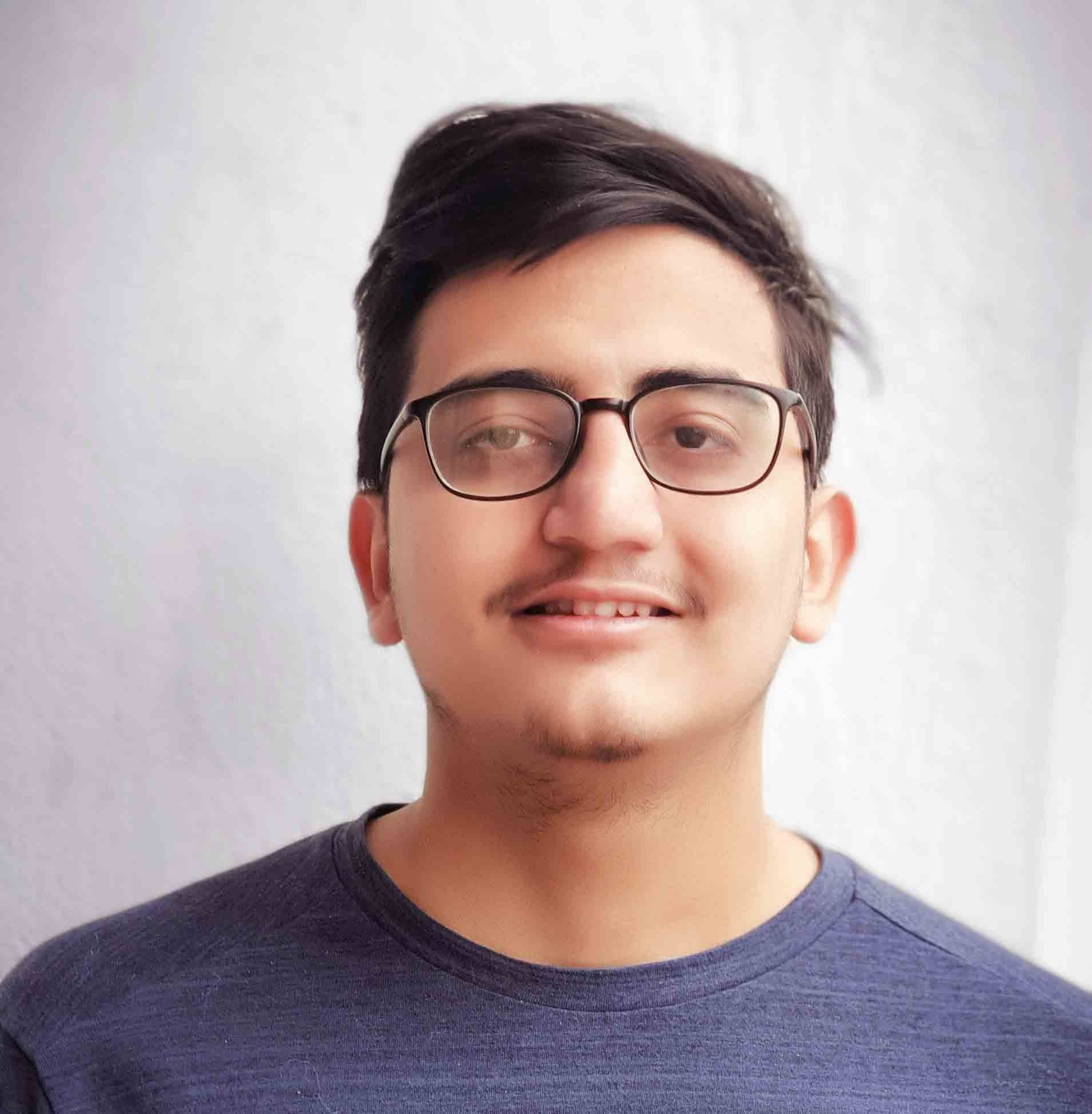}}]{Santosh Pandey}
 is a PhD student in Computer Engineering at Stevens Institute of Technology. He completed his undergraduate from Tribhuvan University, Nepal. His research interests include GPU accelerated high performance computing, machine learning, and graph analytics.
\end{IEEEbiography}
\vspace{-0.5in}  

\begin{IEEEbiography}[{\includegraphics[width=1in,height=1.25in,clip,keepaspectratio]{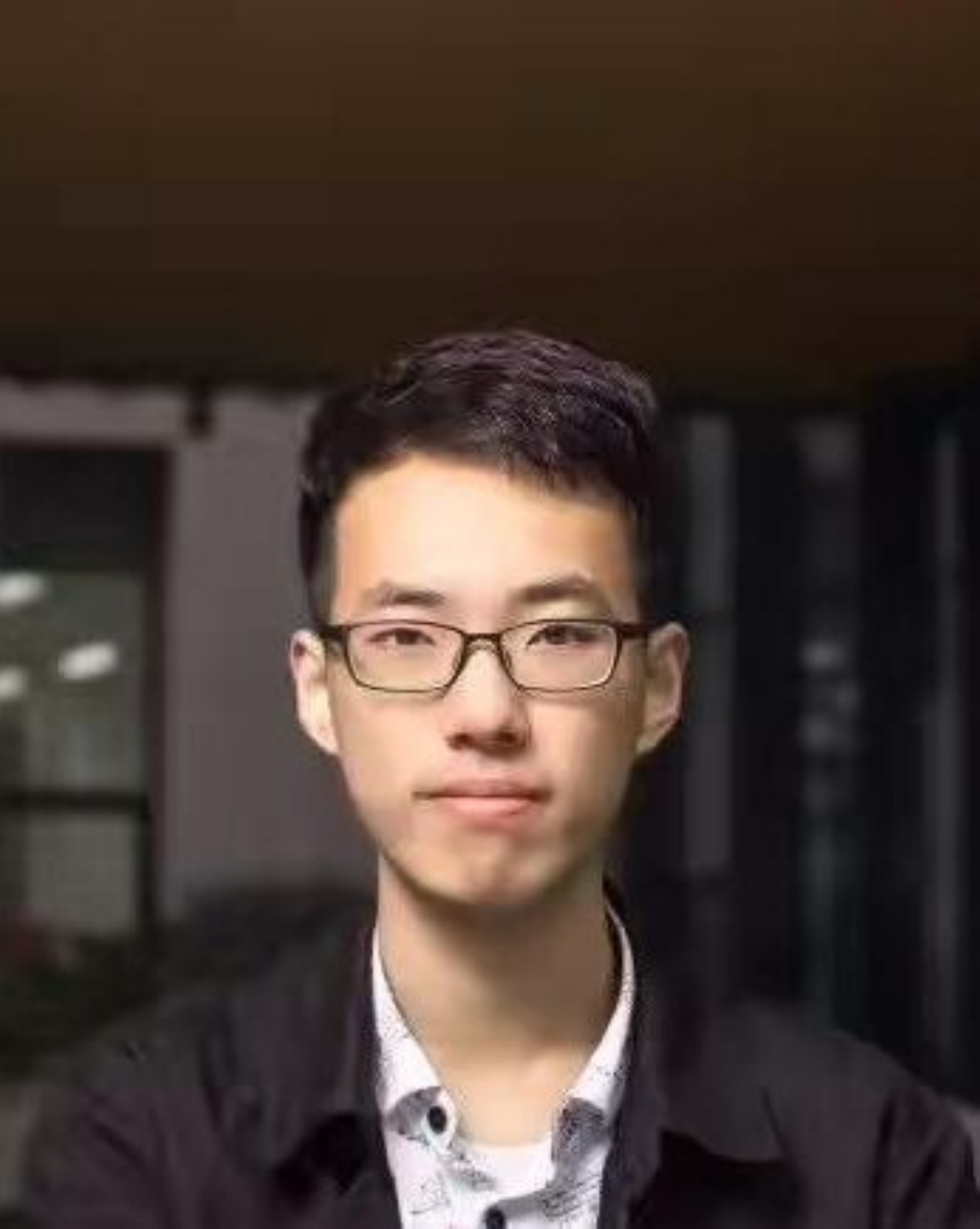}}]{Zhibin Wang}
  received the B.E. degree from the Department of Computer Science and Technology, Nanjing University of Aeronautics and Astronautics in 2018. He is pursuing his Ph.D. degree with the Department of Computer Science and Technology of Nanjing University. His research interests include graph computing, mining and learning.
\end{IEEEbiography}
  \vspace{-0.5in}  

\begin{IEEEbiography}[{\includegraphics[width=1in,height=1.25in,clip,keepaspectratio]{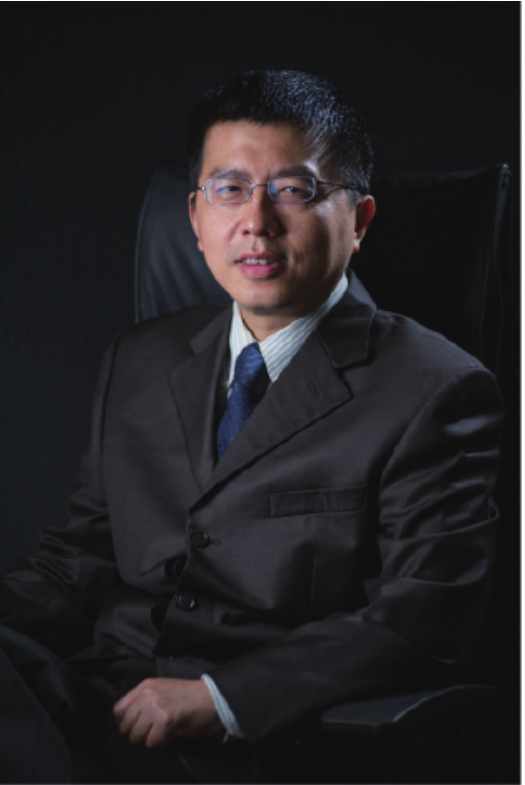}}]{Sheng Zhong}
  received the B.S. and M.S. degrees from Nanjing University in 1996 and 1999, respectively, and the Ph.D. degree from Yale University in 2004, all in computer science. He is interested in security, privacy, and economic incentives.
\end{IEEEbiography}
\vspace{-0.5in}
\begin{IEEEbiography}[{\includegraphics[width=1in,height=1.25in,clip,keepaspectratio]{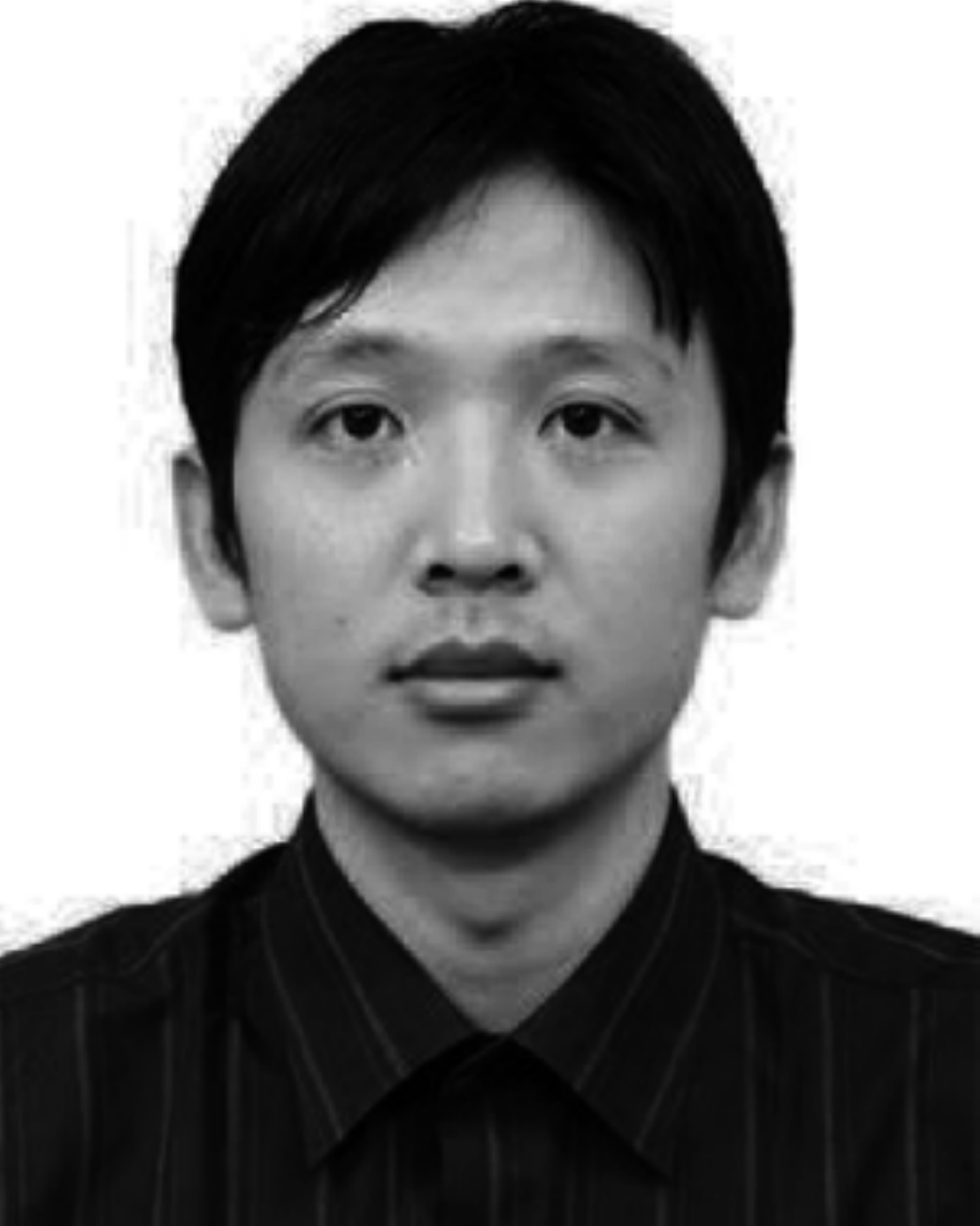}}]{Chen Tian}
  received the BS, MS, and PhD degrees from the Department of Electronics and Information Engineering, Huazhong University of Science and Technology, China, in 2000, 2003, and 2008, respectively. He is an associate professor with the State Key Laboratory for Novel Software Technology, Nanjing University, China. He was previously an associate professor with the School of Electronics Information and Communications, Huazhong University of Science and Technology, China. From 2012 to 2013, he was a postdoctoral researcher with the Department of Computer Science, Yale University. His research interests include data center networks, network function virtualization, distributed systems, Internet streaming, and urban computing.
\end{IEEEbiography}
\vspace{-0.5in}  
\begin{IEEEbiography}[{\includegraphics[width=1in,height=1.25in,clip,keepaspectratio]{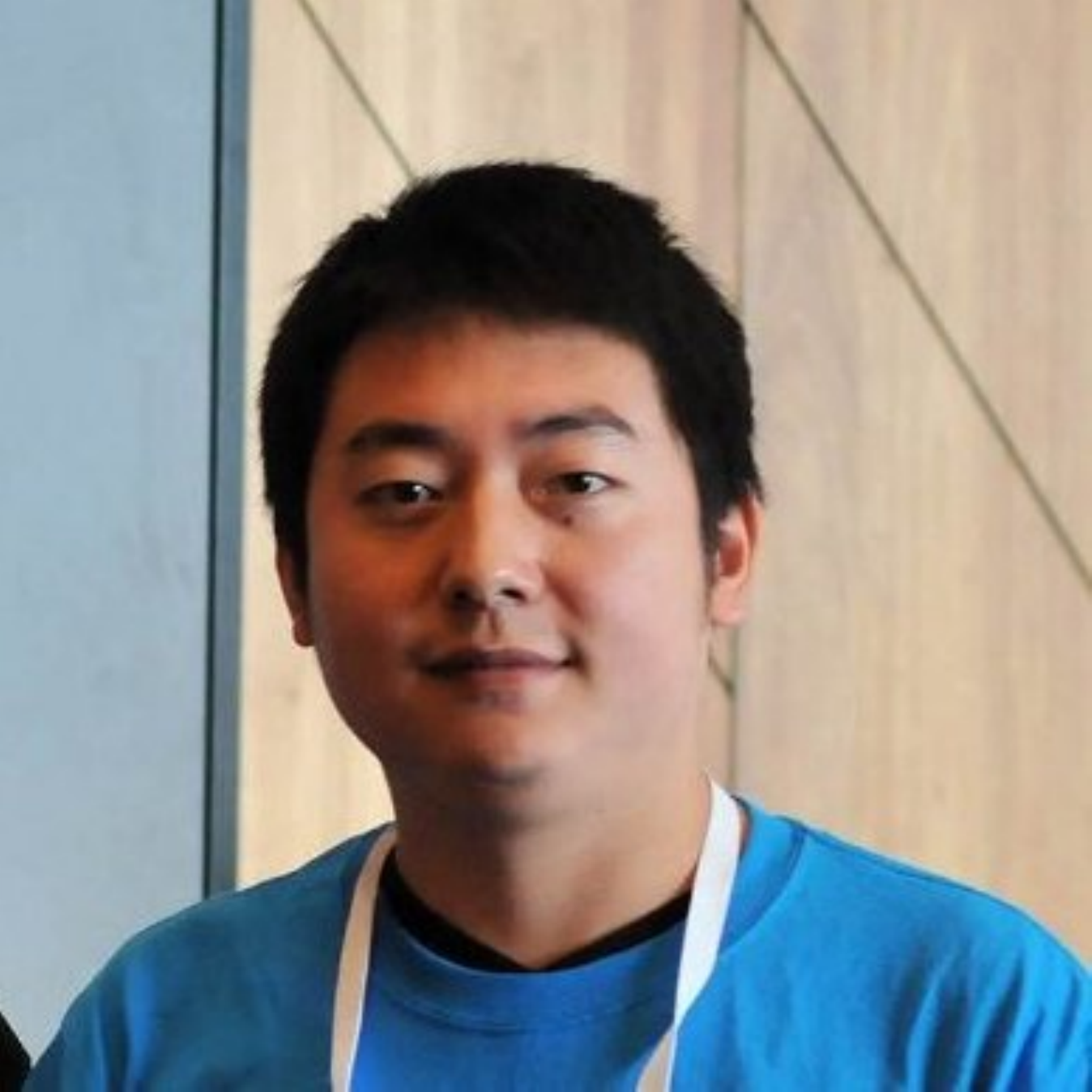}}]{Bolong Zheng}
  received the bachelor's and master's degrees in computer science from the Huazhong University of Science and Technology, in 2011 and 2013, respectively, and the PhD degree from the University of Queensland, in 2017. He is an associate professor with the Huazhong University of Science and Technology (HUST). His research interests include spatio-temporal data management and graph data management.
\end{IEEEbiography}
  \vspace{-0.6in}
\begin{IEEEbiography}[{\includegraphics[width=1in,height=1.25in,clip,keepaspectratio]{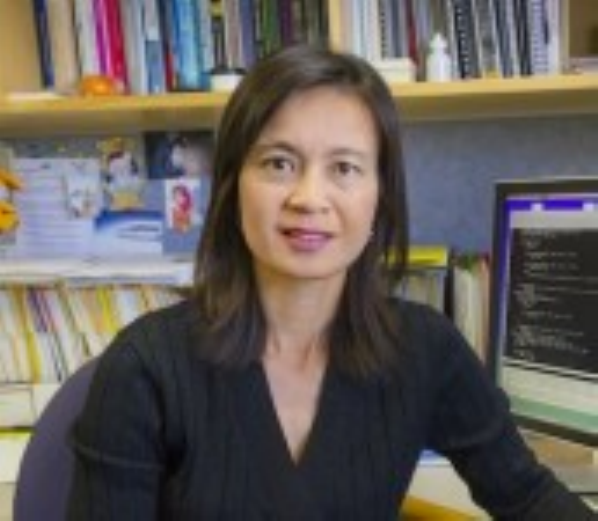}}]{Xiaoye Li}
  is a Senior Scientist in the Computational Research Division, Lawrence Berkeley National Laboratory. She has worked on diverse problems in high performance scientific computations, including parallel computing, sparse matrix computations, high precision arithmetic, and combinatorial scientific computing. She has (co)authored over 120 publications, and contributed to several book chapters. She is the lead developer of SuperLU, a widely-used sparse direct solver, and has contributed to the development of several other mathematical libraries, including ARPREC, LAPACK, PDSLin, STRUMPACK, and XBLAS. She earned Ph.D. in Computer Science from UC Berkeley in 1996. She has served on the editorial boards of the SIAM J. Scientific Comput. and ACM Trans. Math. Software, as well as many program committees of the scientific conferences. She is a SIAM Fellow and an ACM Senior Member.
\end{IEEEbiography}
  \vspace{-0.5in}
\begin{IEEEbiography}[{\includegraphics[width=1in,height=1.25in,clip,keepaspectratio]{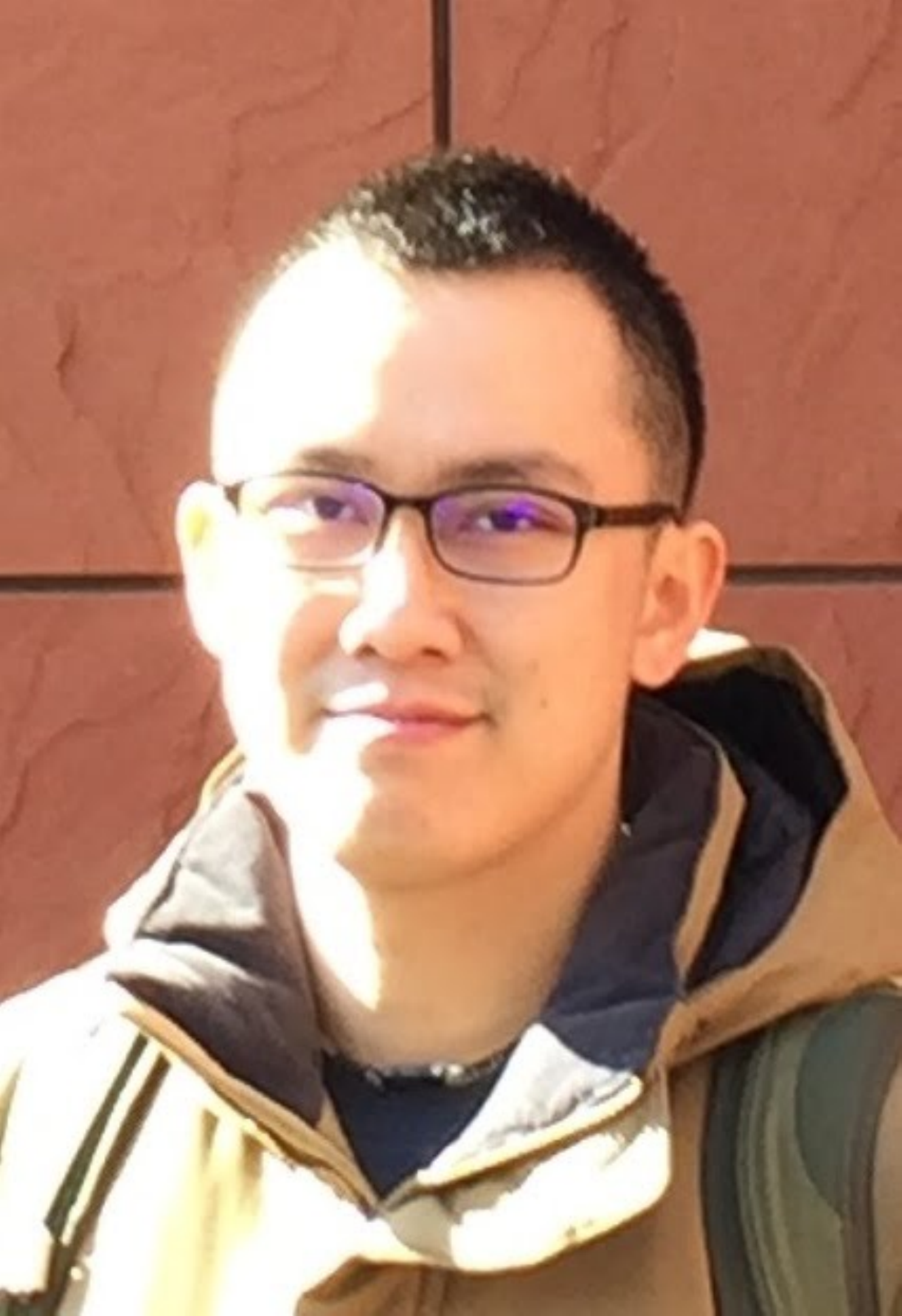}}]{Lingda Li}
  is currently an assistant scientist in the Computational Science Initiative of Brookhaven National Laboratory. He mainly works with Adolfy Hoisie and Barbara Chapman. His current research focuses on memory systems, with an emphasis on architecture simulation, programming models, and compiler.
  Before joining BNL, he worked at the Department of Computer Science of Rutgers University as a Postdoc with Professor Eddy Z. Zhang to carry out GPGPU research between 2014 and 2016, he obtained his PhD degree in computer architecture from the Microprocessor Research and Development Center, Peking University in 2014
\end{IEEEbiography}
  \vspace{-0.5in}
\begin{IEEEbiography}[{\includegraphics[width=1in,height=1.25in,clip,keepaspectratio]{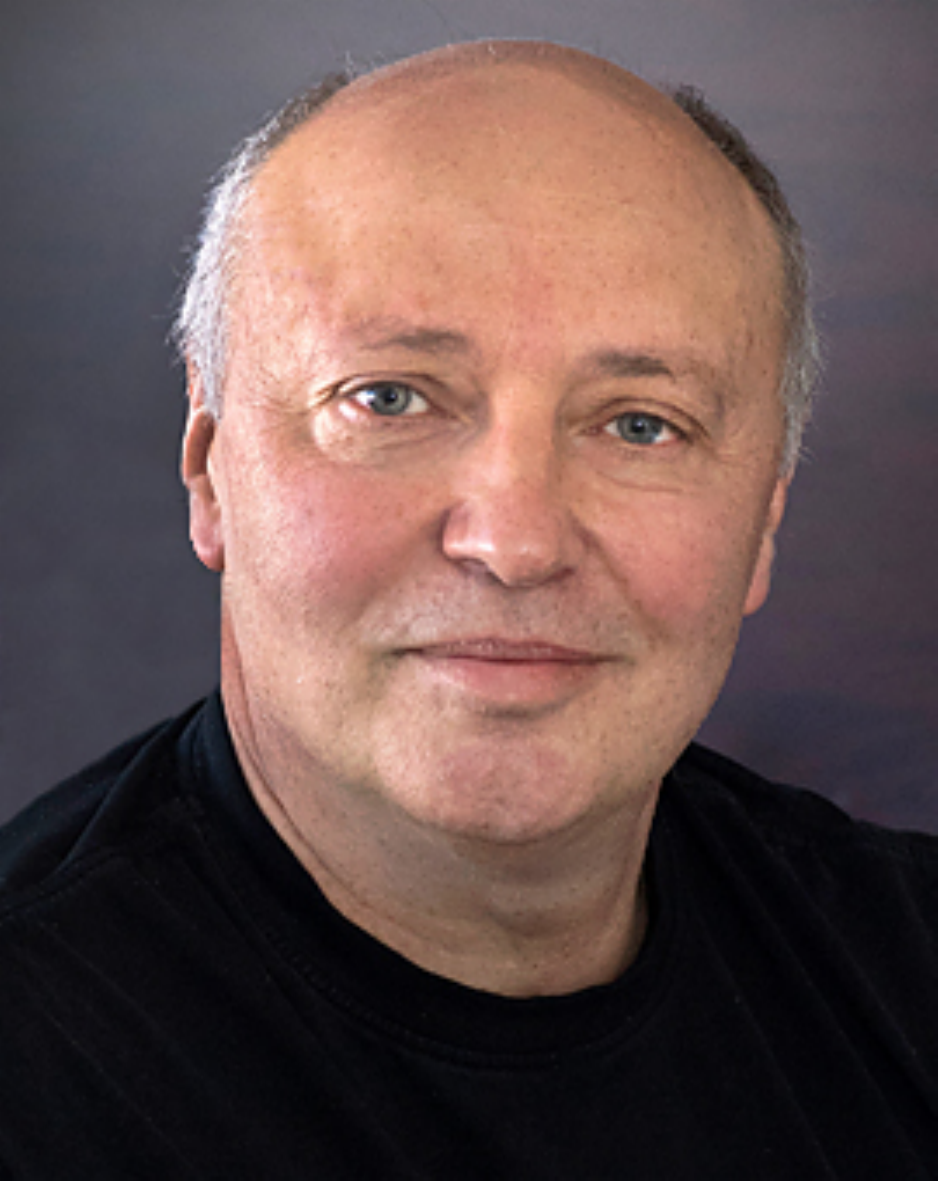}}]{Adolfy Hoisie}
  is a Laboratory Fellow and chief scientist for computing at the Pacific Northwest National Laboratory, where he also directs multiple large-scale projects and research groups. His research focuses on performance and power analysis and modeling of extreme-scale systems, applications, and architectures. Among other awards for research and teaching excellence, Hoisie received the Gordon Bell Award in 1996 for his leading-edge work in parallel computing. He received a PhD in computational science from Cornell University.
\end{IEEEbiography}
  \vspace{-0.5in}
\begin{IEEEbiography}[{\includegraphics[width=1in,height=1.25in,clip,keepaspectratio]{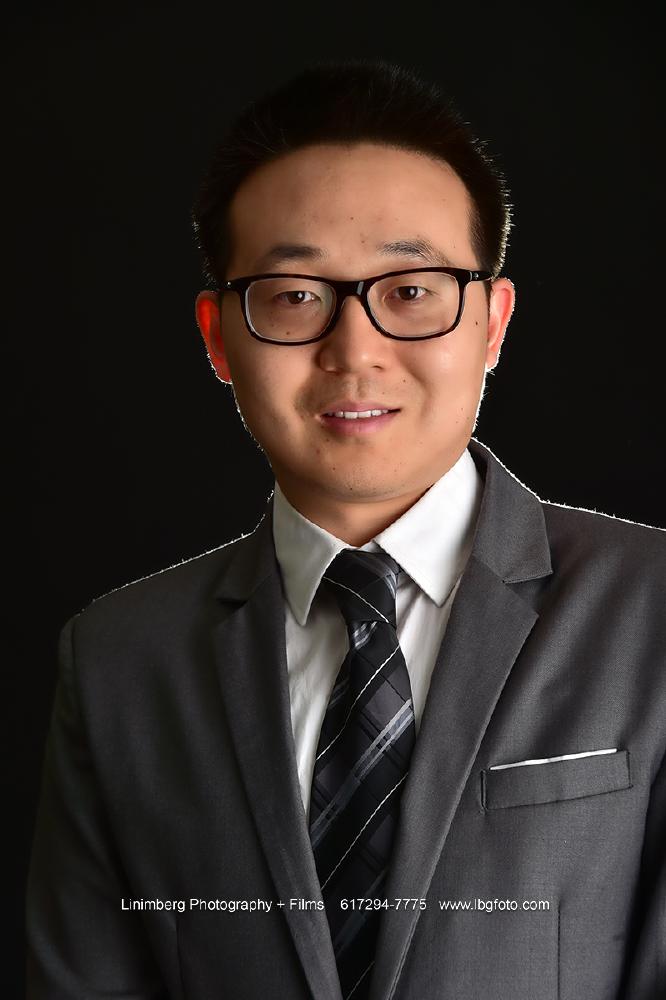}}]{Caiwen Ding}
  is an assistant professor in the Department of Computer Science \& Engineering at the University of Connecticut. He received his Ph.D. degree from Northeastern University (NEU) in 2019, supervised by Prof. Yanzhi Wang.
His interests include Machine Learning \& Deep Neural Network Systems; Computer Vision, Natural Language Processing; Computer Architecture and Heterogeneous Computing (CPUs/FPGAs/GPUs); Non-von Neumann Computing and Neuromorphic Computing;
His work has been published in high-impact conferences (e.g., AAAI, ASPLOS, ISCA, MICRO, HPCA, FPGA, DAC, DATE, ISLPED). His work on Block-Circulant Matrix-based Smartphone Acceleration has received the Best Paper Award Nomination at DATE 2018.
\end{IEEEbiography}
\vspace{-0.3in}
\begin{IEEEbiography}[{\includegraphics[width=1in,height=1.25in,clip,keepaspectratio]{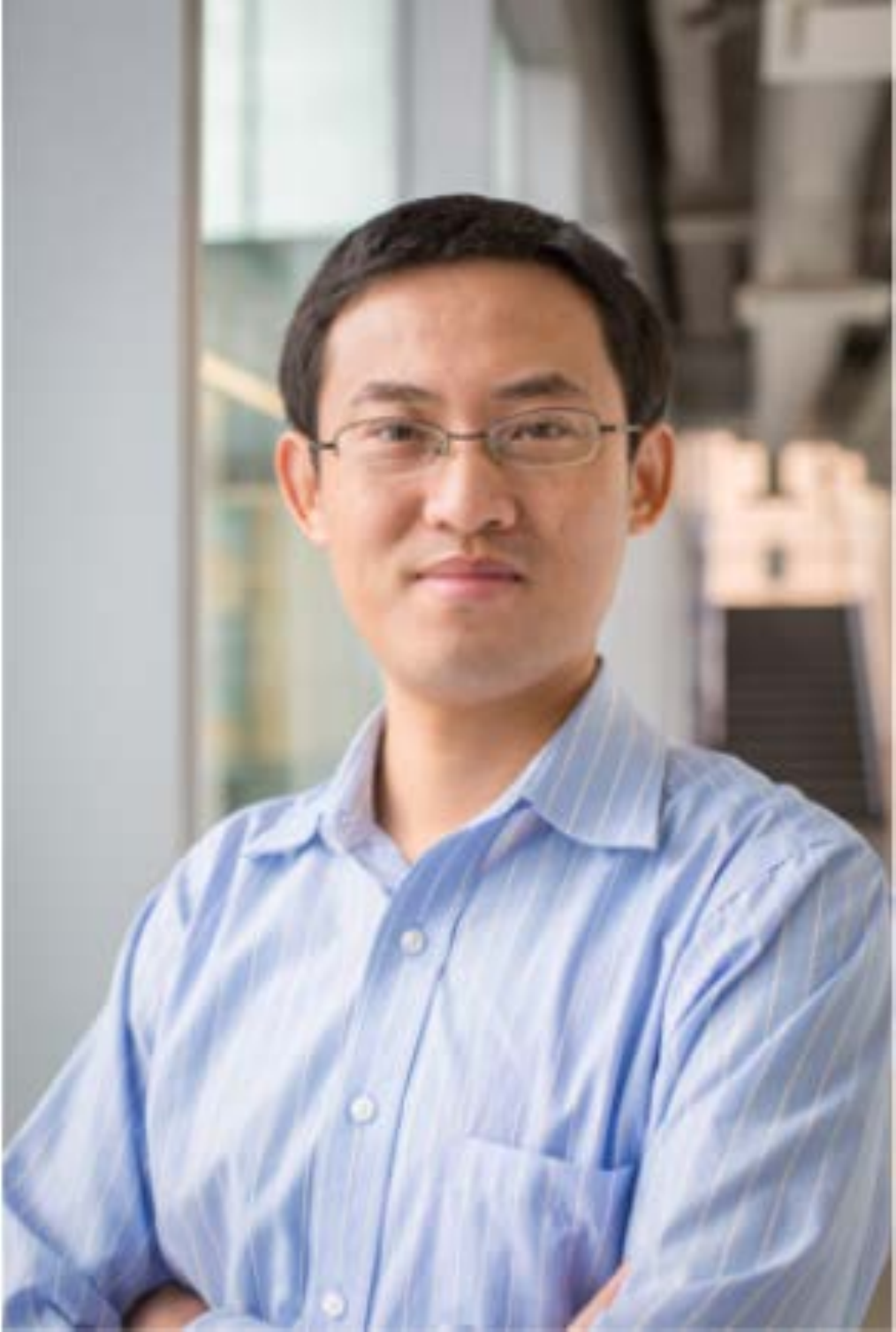}}]{Dong Li}
  is an assistant professor in the Department of Electrical Engineering and Computer Science at the University of California, Merced. His research interests include high-performance computing, performance modeling, programming models, architecture, and runtime. In collaboration with industry, DOE laboratories, and universities, Dong has been involved in various DOE, DOD and NSF projects related to HPC. His work was published in a number of premier HPC conferences. He is a program committee member in a number of international conferences and workshops. Li has a PhD in computer science from Virginia Tech. 
\end{IEEEbiography}

\begin{IEEEbiography}[{\includegraphics[width=1in,height=1.25in,clip,keepaspectratio]{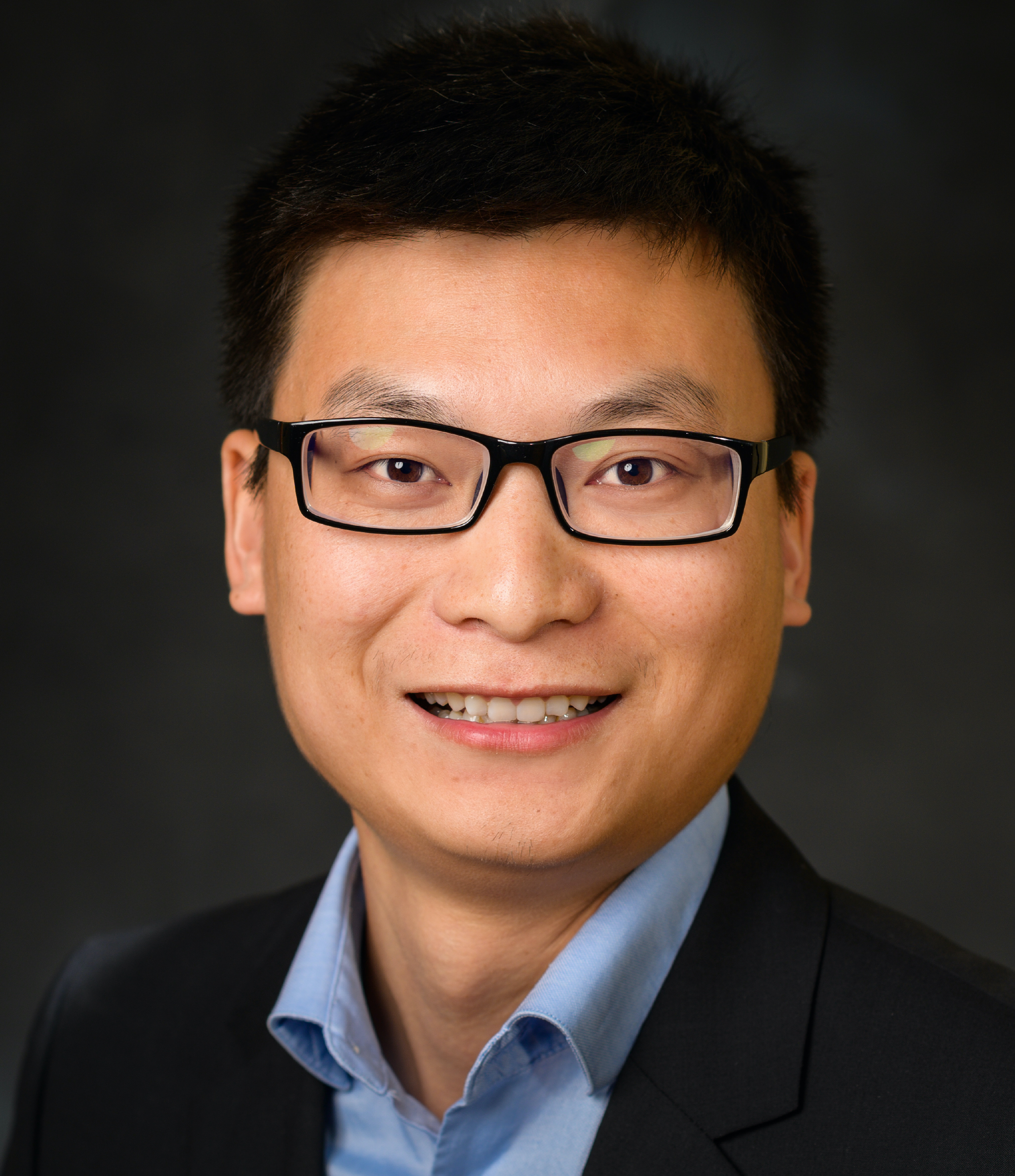}}]{Hang Liu} is an Assistant Professor of Electrical and Computer Engineering at Stevens Institute of Technology. Prior to joining Stevens, he was an assistant professor at the Electrical and Computer Engineering Department of University of Massachusetts Lowell. He is on the editorial board for Journal of BigData: Theory and Practice, a program committee member for SC, HPDC, and IPDPS, and regular reviewer for TPDS and TC. He earned his Ph.D. degree from the George Washington University 2017. He is the Champion of the MIT/Amazon GraphChallenge 2018 and 2019, and one of the best papers awardee in VLDB '20. 
 
\end{IEEEbiography}

\clearpage

\onecolumn
\newpage

\end{document}